\documentclass[preprint]{revtex4}
\usepackage{graphicx}

\usepackage{amssymb,amsfonts,amsmath,color}

\begin{document}

\title{A comprehensive phase diagram for logistic populations in fluctuating environment.   }

\author{Yitzhak Yahalom, Bnaya Steinmetz and Nadav M. Shnerb}

\affiliation{Department of Physics, Bar-Ilan University,
Ramat-Gan IL52900, Israel.}

\begin{abstract}
\noindent
Population dynamics reflects an underlying birth-death process, where the rates associated with different events may depend on external environmental conditions and on the population density. A whole family of  simple and popular deterministic models (like logistic growth) support a transcritical bifurcation point between an extinction phase and an active phase. Here we provide a comprehensive analysis of the phases of that system, taking into account both the endogenous demographic noise (random  birth and death events) and the effect of environmental stochasticity that causes variations in birth and death rates. Three phases are identified: in the  inactive phase the mean time to extinction $T$  is independent of the carrying capacity $N$, and scales logarithmically with the initial population size. In the  power-law phase  $T \sim N^q$ and the exponential phase  $T \sim exp(\alpha N)$. All three phases and the transitions between them are studied in detail. The breakdown of the continuum approximation is identified inside the power-law phase, and the accompanied changes in decline modes are analyzed. The applicability of the emerging picture to the analysis of ecological timeseries and to the management of conservation efforts is briefly discussed.
\end{abstract}

\maketitle

\section{Introduction}

All environments fluctuate. Temperature, precipitation, wind velocity, predation pressure and food availability vary on all relevant spatio-temporal scales, from microns to continents, from microseconds to ages. These fluctuations affect the reproductive success of individuals and this, in turn, yields abundance variations that govern community structure and the evolutionary process. Through this paper we consider the fate of a  population in stochastic environment.

The effect of environmental variations on conspecific individuals may be classified according to the level of correlations. In one extreme we think about an accidental encounter with a predator or with a piece of food, events that affect individuals in an uncorrelated manner. In the other extreme, droughts or cold waves may affect coherently entire populations.       In reality one should expect a whole spectrum of stochastic perturbations and disturbances that influence groups of variable size. Nevertheless, for the sake of simplicity the corresponding theory distinguishes between \emph{demographic stochasticity} (aka drift, shot noise), i.e., those aspects of noise that influence individuals in a completely uncorrelated manner, and \emph{temporal environmental stochasticity}, that acts on entire populations \cite{lande2003stochastic}.

Demographic noise (genetic or ecological drift) yields abundance fluctuations that scale  with the square root of the population size, while environmental stochasticity leads to variations that scale linearly with the abundance. Accordingly, one should expect that environmental stochasticity is the dominant mechanism . A few recent large-scale empirical studies show that abundance fluctuations for populations with $n$ individuals indeed scale linearly with $n$~\cite{leigh2007neutral,kalyuzhny2014niche,kalyuzhny2014temporal,
chisholm2014temporal}. On the other hand, demographic stochasticity provides the only scale against which the intensity of environmental variations may be measured~\cite{kessler2014neutral,hidalgo2017species,danino2016effect}. Moreover, since demographic noise controls the low-density states of the system, it dictates important quantities like extinction times and species richness~\cite{danino2018theory}.  Consequently, the study  of models that combine deterministic effects,  temporal environmental stochasticity and demographic noise,  received a considerable  attention during the last years~\cite{kessler2014neutral,kessler2015neutral,saether2015concept,cvijovic2015fate,
kalyuzhny2015neutral,danino2016stability,
fung2016reproducing,hidalgo2017species,wienand2017evolution}.

Here we would like to consider, within this framework, the simplest and the most important model of population dynamics, in which the deterministic evolution of the abundance $n$  is logistic or logistic-like. Some of our main findings were presented in brief in a recent work \cite{yahalom2018phase}; in this paper we provide the full analysis and discuss in detail the various transitions in that system and the implications of our work to the theory of population and community dynamics.

The logistic equation,
\begin{equation} \label{eq1}
\frac{dn}{dt} = r_0 n -\beta n^2,
\end{equation}
describes a very simple process that involves exponential (Malthusian) growth and negative density response (usually due to resource depletion). $r_0$ corresponds to the low-density growth rate of the population. As $n$ increases, the growth decreases until the population saturates at $n^* = r_0/\beta$.

Technically speaking, the deterministic dynamics of Eq. (\ref{eq1}) supports a transcritical bifurcation. When $r_0$ is positive the fixed point at $n = 0$  is unstable and $n^*$ is a stable fixed point. If $r_0<0$ the only stable and feasible fixed point is the extinction state $n = 0$.  At the transition, $r_0=0$, the population decays asymptotically like $1/t$, as opposed to the exponential decay below the transition.

A wide variety of population dynamics models support such a transcritical bifurcation.  These include the $\theta$-logistic  equation (where  ${\dot n} = r_0 n [1-( n/K)^\theta]$, The logistic system corresponds to $\theta=1$),  ceiling models (growth rate is kept fixed but the population cannot grow above a given carrying capacity, corresponds to $\theta = \infty$), Ricker dynamics and so on. For the sake of concreteness, in what follows we will analyze a specific model. However, in section \ref{universality} we will show that the outcomes of our study hold for all the systems that belong to the transcritical bifurcation class.

Since the actual number of individuals in a population is always an integer,  Eq. (\ref{eq1}) and its variants can only be understood as the deterministic limit of an underlying stochastic process, in  which birth ($A \to  2A$), death ($A \to \varnothing$) or competition (say, $A+A \to A$) occur at random. For such a process the empty state $n=0$ is the only absorbing state, so each population, for any set of parameters, must reach extinction in the long run. Under pure demographic stochasticity, when each individual is affected independently by the environmental fluctuations, the sign of  $r_0$ determines the mean time to extinction $T$.  When $r_0<0$,   $T$ is logarithmic in the initial population size $n$ and does not depend on the carrying capacity.  For $r_0>0$ the mean time  $T$ grows exponentially with  $n^*$. At the bifurcation transition point ($r_0=0$) the functional form of the time to extinction depends on the initial condition $n$ as we shall see below (section \ref{hom}).

Another aspect of the transition between the logarithmic and the exponential behavior has to do with the applicability of the corresponding continuum  (Fokker-Planck or backward Kolmogorov) equations. These differential equations emerge from the underlying difference (master) equation of the stochastic process via the continuum approximation, which fails when the relevant function [e.g., the mean time to extinction given $n$, $T(n)$] is not smooth enough over the integers. The continuum approximation fails in the exponential phase, and gives wrong estimations for the lifetime of the system. To overcome this difficulty, a WKB technique has been proposed by Kessler and Shnerb~\cite{kessler2007extinction}, and we will implement a similar approach in the relevant cases hereon.

The aim of this paper is to provide a comprehensive analysis of a logistic system that supports a finite number, $N$, of individuals ($n^*$ is proportional to $N$) under the influence of  both demographic and environmental stochasticity. This problem was considered by a few authors \cite{lande2003stochastic,kamenev2008colored,spanio2017impact,wada2018extinction} for the case where the strength of  environmental fluctuations is \emph{unbounded}, for example when  the amplitude of these variations is an Ornstein-Uhlenbeck process. In that case there are always (very rare) periods in which the net growth rate is negative, and (as we shall see below) these periods dominate the large $N$ asymptotic behavior of  extinction times. As a result, the system admits only two phases: an inactive (logarithmic) phase and a power law phase~\cite{vazquez2011temporal}, but there is no exponential phase.

We consider a system under dichotomous (telegraphic) noise, with finite amplitude $\sigma$ and correlation time $\tau$. Since the noise is bounded, above a certain value of $r_0$ the growth rate is always positive, so the system allows for   a phase in which $T$ growth exponentially with $N$, and for a transition that has not been explored yet, between the power-law phase and the exponential phase. The continuum (diffusion) approximation used in former studies breaks down inside the power-law phase. To study the deep power-law region and the transition to exponential behavior we  developed a WKB technique which is shown to yield the correct results.  This allows us to provide a comprehensive analysis of all the three phases and the  transitions or crossovers between them.

Inside the power-law phase we identify three (perhaps related) transitions, or crossovers:  the failure of the continuum approximation, the opening of a spectral gap for the corresponding Markov matrix and a qualitative shift between soft and sharp decline modes. In what follows we shall discuss these transitions and their relevance to the analysis of empirical datasets.

In the next section a generic phase diagram for transcritical systems with bounded environmental stochasticity is presented  and discussed. The reader is referred, from each part of the diagram, to the relevant section.

\section{The phase diagram}

\begin{figure}
\includegraphics[width=8cm]{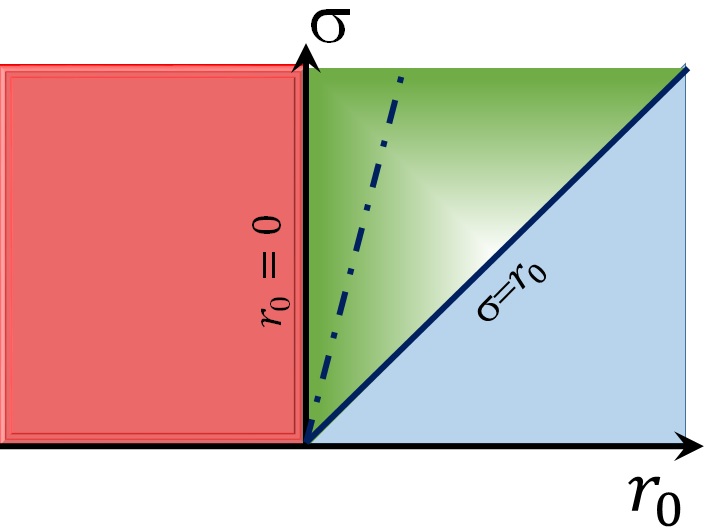}
\caption{A phase diagram for a logistic system under demographic and bounded environmental stochasticity, presented in the $r_0$-$\sigma$ plane. In the inactive phase  ($r_0  <0$, red) the time to extinction scales like $\ln n(t=0)$ where $N$ plays no role, except of setting an upper bound for $n(t=0)$. In the exponential phase  $r_0>\sigma$ (blue) the extinction time grows exponentially with $N$.  When $\sigma>0$ the logarithmic and the exponential phases are separated by a finite power-law region (green). At the logarithmic-power transition ($r_0=0$, $\sigma>0$) $T$ grows like $\ln^2 N$, while at the power-exponential transition T is a stretched exponential in N.  The dashed-dotted line indicates the failure of the continuum (diffusive) approximation, which is correlated with the transition between soft and sharp decline mode.  }\label{fig1}
\end{figure}

This paper is organized around the phase diagram shown in  Figure \ref{fig1}. The $x$-axis of figure \ref{fig1} is the time averaged value of the linear growth rate, $r_0$, and the $y$ axis corresponds to $\sigma$, the amplitude of environmental variations. Demographic noise affect the system in any case.

Three phases are identified:

\begin{itemize}
  \item \textbf{Extinction (logarithmoc) phase:}  In the red region, $r_0 <0$, the time average of the linear growth rate is negative. In this regime the time to extinction grows  \emph{logarithmically} with the \emph{initial population size}, $T \sim \ln n$.  The overall capacity of the system, $N$, only limits the maximum value of $n$ at $t=0$ so it sets the scale for the maximum lifetime, but for fixed $n(t=0)$, the mean lifetime $T$ is independent of $N$.

      In this logarithmic regime one may analyze the corresponding stochastic process by taking the continuum limit of the backward Kolmogorov equation, as explained in Appendix \ref{deriv}. Results for this regime are presented in Section \ref{contres}.

  \item \textbf{Power-law phase:} In the green region where $0<r_0<\sigma$, the mean growth rate is positive but the instantaneous growth rate may become negative because of the environmental variations. Below (sections \ref{contres} and \ref{WKBsec}) we consider this regime and show that the time to extinction grows like a power-law in $N$. Here the large-$N$-dependence of $T$ is not affected by the initial conditions and the difference between $T$ for a single individual and for $N$ individuals appears only in the prefactor. Put it another way, the chance of establishment for a single individual is $N$ independent.

  \item \textbf{Active (exponential) phase:} In the blue region $r_0>\sigma$, the linear growth rate is always positive. In this regime the mean time to extinction grows exponentially with $N$, $T \sim \exp(\alpha N)$. In section \ref{expsec} we show that $\alpha$ is related to the dynamics of a system with time-independent  growth rate $r_0 - \sigma$, while growth rate variations contribute only to the power-law pre-factor of this exponent.
\end{itemize}

Three types of transition regions between these  phases are discussed below:

\begin{itemize}
  \item  When there are no environmental variations ($\sigma =0$), the transition point at the origin of  the $x$-axis separates the logarithmic ($N$-independent) and the exponential phase. This point is analyzed in section \ref{hom}. At the transition point the scaling of $T$ with $N$ depend on the initial conditions, and ranges between $T \sim \ln N$ for a single individual to $T \sim \sqrt{N}$ when the initial state is a finite fraction of $N$ (Eq. \ref{rescrit0}).

  \item  The $y$ axis of Fig. \ref{fig1}, where $r_0 = 0$ but $\sigma >0$, marks the transition between the logarithmic and the power-law  phase when the environment fluctuates randomly. Again, along this line the $N$ scaling of the time to extinction depends on the initial conditions, but now it runs between $\ln N$ for a single individual to $\ln^2 N$ for finite fraction. See discussion in Section \ref{critical}.

  \item  The  line $\sigma = r_0$ marks the transition between the power-law and the exponential phase. This transition line is characterized by a stretched exponential scaling and is discussed in Section \ref{expsec}.
 \end{itemize}

The phase diagram is based on the features of the mean time to extinction $T$. In section \ref{pdf} we present some considerations and results for
a more general quantity, the probability distribution function for exit times, $f(t) \  dt$.

 As mentioned above, the continuum approximation breaks down inside the power-law phase, and this is indicated by the \textbf{dashed-dotted line} in Fig. \ref{fig1}. In section \ref{WKBsec} we present a WKB analysis which is valid even where the continuum approximation breaks and converges to the continuum result when $r_0$ is much smaller than $\sigma$. The power-law exponent predicted by this WKB analysis diverges when $\sigma \to r_0$, marking the transition to the third, exponential regime.

 Another transitions that appears inside the power-law phase, and are probably related to the breakdown of the continuum approximation, are the emergence of a spectral gap for the corresponding Markov matrix, the crossover between soft and sharp decline modes and the qualitative alteration of the quasi-stationary probability distribution function, these phenomena are discussed in Sections \ref{WKBsec} and \ref{pdf}.

In the next sections we present and analyzed a specific microscopic model but, as mentioned above, we believe that the general picture emerges -  the phases, the different functional dependencies of $T$ on $N$, and the characteristics of the transitions - are generic and will characterize any system that supports a transcritical bifurcation under demographic noise and bounded environmental stochasticity. In Section \ref{universality} we implement our WKB analysis to support this argument.

Finally, the emerging insights and their relevance to the theoretical understanding and to the empirical analysis of many practical problems, ranging from the assessment of population viability and the management of conservation efforts to the general theory of species coexistence. A preliminary discussion of these points is presented in Section \ref{practical}.

\section{The model and the continuum (diffusion) approximation} \label{model}

The logistic equation (\ref{eq1}) is the deterministic limit of many underlying stochastic processes. We would like to compare our analytic results with the numerical solutions of the corresponding Markov process, thus we prefer a process in which the total number of individuals is bounded. To that aim, we use a genetic model for two-allele (species) competition with one sided mutation, as defined in \cite{karlin1981second}. One-sided mutation ensures that the system supports only one absorbing state, that we define as the extinction state of the focal species. In section \ref{universality} below we explain why the qualitative characteristics of the phase space diagram do not depend on the microscopic features of the model.

 Let us consider  a system with $N$ individuals, $n$ of them belong to species A and $N-n$ to species B. At each elementary step two individuals are drawn at random for a duel, the loser dies and the winner produces a single offspring.  This is a zero sum model so $N$ is kept fixed and there is only one degree of freedom, $n$. The endogenous demographic stochasticity is related to the discrete nature of the state variable $n$.

 In case of an intraspecific competition (if two A-s or two B-s were chosen) both individuals have the same relative fitness and each of them wins with probability $1/2$. When interspecific duel takes place the chance of the $A$ individual to win is
\begin{equation}
P_A = \frac{1}{2}+\frac{s(t)}{4},
\end{equation}
where the dependence of $s$ on time reflects the effect of environmental fluctuations.

The outcome of each elementary duel is birth and death: the loser dies, and the winner produces a single offspring. When an A individual wins a  duel, the offspring is also an A with probability $1-\nu$, and with probability $\nu$ the offspring is a $B$-individual. One may consider that as a mutation from A to B. On the other hand, when a B wins a duel the offspring is always B, so the mutation is one-sided~\cite{karlin1981second}. The possible outcomes of all kinds of duels are (note that the expressions above the arrows are probabilities, not rates),
 \begin{eqnarray} \label{eq1a}
 B+B \xrightarrow{1} 2B \qquad A&+&A  \xrightarrow{1-\nu} 2A  \qquad A+A  \xrightarrow{\nu} A+B \nonumber \\
 A+B \xrightarrow{1-P_{A}} 2B \qquad A&+&B \xrightarrow{P_{A}(1-\nu)} 2A \qquad A+B \xrightarrow{\nu P_{A}} A+B.
 \end{eqnarray}

 \begin{figure}[h]
\includegraphics[width=8cm]{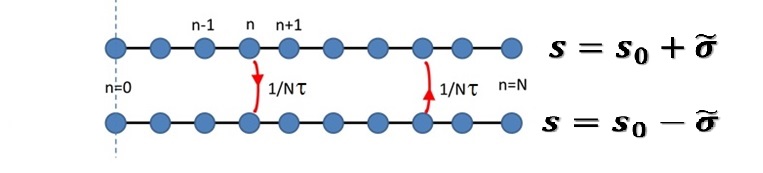}
\caption{A sketch of the possible states of two-allele competition with one sided mutation, demographic stochasticity and two environmental states. The abundance of the focal species is $n$, and in each elementary duel it may change to either $n+1$ or $n-1$. Under dichotomous noise $s = s_0 + \tilde{\sigma}$ (upper channel) or $s = s_0 - \tilde{\sigma}$ (lower  channel). The chance of environmental flip is $1/N\tau$ per elementary step, so the persistence time of the environment is $\tau$ generations, where a generation is $N$ elementary duels. The state $n=0$ is absorbing, but the system may escape from the state $n=N$ due to mutations.  }\label{system}
\end{figure}

To specify the stochastic process completely we have to define the dynamics of $s(t)$. We define $s(t) = s_0 + \eta(t)$, where $s_0$ is the mean (logarithmic) relative fitness of species A and $\eta(t)$ is a zero-mean stochastic process. As we shall see, in many cases the only important characteristics of this process are its correlation time $\tau$, its mean $s_0$ and its amplitude $\sigma$.

We consider a system with dichotomous (telegraphic) environmental noise, so $\eta = \pm \tilde{\sigma}$. After each elementary step $\eta$ may switch (from $\pm \tilde{\sigma}$ to $\mp \tilde{\sigma}$, see Fig. \ref{system}) with probability $1/N\tau$, so the persistence time of the environment is taken from a geometric distribution with mean $N \tau$ steps, or (if time is measured in units of generation, $N$ duels correspond to a single generation) $\tau$ generations. Dichotomous noise of this kind may imitate the features of other generic distributions, like Gaussian or Poisson noise, see Appendix \ref{telegraphic}. However, dichotomous noise is bounded, and this makes a difference in some circumstances, as explained in section \ref{WKBsec}.

The chance of an interspecific duel for two, randomly picked individuals is $$F_n = 2n(N-n)/N (N-1),$$ and the chance for an A-intraspecific duel is $$Q_n = n(n-1)/N(N-1).$$
 We define $$P_A^+ =1/2 + s_0/4 + \tilde{\sigma}/4$$  as the chance of an A individual to win against a B in the $+
\tilde{\sigma}$ state, and $$ P_A^-=1/2 + s_0/4 - \tilde{\sigma}/4$$ as its chance to win in the minus state.

In analogy to Eq. (\ref{eq1}), in our  model the  dynamics of $n$ in the limit $N \to \infty$ (without demographic stochasticity)   is given by,
\begin{equation} \label{deter}
{\dot n} = r(t) n -\beta(t) n^2,
\end{equation}
where $$r(t) \equiv s_{eff} \pm \sigma - \nu,$$ $$\beta(t) = N \left(s_{eff} \pm \sigma\right).$$  Here $s_{eff} \equiv s_0(1-\nu/2)$ and  $\sigma \equiv \tilde{\sigma} (1-\nu/2)$.  The $\pm$ correspond to the plus and the minus state.

In Appendix \ref{deriv} we provide the derivation of a single, second order differential equation for the mean time to extinction starting from frequency $x \equiv n/N$, where the mean is taken over both histories and initial environmental conditions. This equation is,
\begin{equation}\label{eq9}
\left(s_{eff} - \frac{\nu}{1-x} + g (1-2x) \right)T'(x) + \left(\frac{1}{N} + g x(1-x) \right) T''(x)  = -\frac{1}{x(1-x)},
\end{equation}
and the boundary conditions are $$ T(0) = 0 \qquad T'(1) = \frac{1}{\nu}.$$
The constant,
\begin{equation}
g \equiv \frac{\sigma^2 \tau}{2},
\end{equation}
is the strength of the environmental fluctuations. Under pure environmental stochasticity the growth/decay of $\ln x_0$ during $\tau$ is  $\ln x_0 \to \ln x_0 \pm \sigma \tau$, so $g$ is the diffusion constant of the system along the logarithmic abundance axis.

The derivation of Eq. \ref{eq9} is presented in Appendix \ref{deriv} using two approaches. The first is based on our specific model and the second implements a generic Moran process~\cite{moran1962statistical,ewens2012mathematical}. The numerical results presented below are based on the linear discrete backward Kolmogorov equation (\ref{eqb1}). The numerical techniques we have used (matrix inversion, a transfer matrix approach) are explained in detail in Appendix \ref{methods}. In what follows we compare our analytic results with the outcomes of these numerical calculations, where the use of algorithms for sparse matrices and quadruple precision  allows us to consider systems up to $N = 10^7$.

The two alternative derivations presented  in Appendix \ref{deriv} are based on the continuum approximation, i.e., on replacing $T_{n+1}$, say, by $T(x)+T'/N +T''/2N^2$. Below (section \ref{WKBsec}) we will discuss possible failures of that approximation and will explain how to overcome these difficulties, but in the next few sections we assume the validity of  Eqs. (\ref{eq9}) and track its implications.

\section{Mean time to extinction: an asymptotic matching approach} \label{contres}

To find the large $N$ behavior of the lifetime $T(x)$ we would like to solve  Eq. (\ref{eq9}) in different regimes, then we will use an asymptotic matching techniques to determine the constants of  integration.

In the \emph{inner regime} $x \ll 1$ so $1-x \sim 1$. Accordingly,  (\ref{eq9}) takes the form
\begin{equation}\label{eq11}
[(1+Gx)T_{in}']'+r_0 N T_{in}' =\frac{-N}{x},
\end{equation}
with $G \equiv Ng$ and the relevant  boundary condition is $T(0) = 0$.
One integration is thus trivial, and using an integrating factor one obtains,
\begin{equation}\label{eq26b}
T_{in}(x) = \frac{c_1}{Nr_0}\left( 1-\frac{1}{(1+Gx)^{r_0/g}} \right)- \frac{N}{(1+Gx)^{r_0/g}}I(x),
\end{equation}
where,
\begin{equation} \label{Idef}
I(x) \equiv \int_0^x \frac{\ln t \  dt}{(1+Gt)^{1-r_0/g}}.
\end{equation}

In the \emph{outer regime} $x \gg 1/G$, the demographic stochasticity term $1/N$ in Eq. (\ref{eq9}) is negligible. This yields,
\begin{equation}\label{eq27}
[gx(1-x)]T_{out}''(x)+\left(s_{eff} + g(1-2x)-\frac{\nu}{1-x}\right)T_{out}'=\frac{-1}{x(1-x)},
\end{equation}
or,
\begin{equation}\label{eq27a}
Q'(x)+\left(\frac{s_{eff}/g}{x(1-x)} -\frac{\nu/g}{x(1-x)^2}\right)Q(x)=\frac{-1}{x(1-x)},
\end{equation}
with  $Q \equiv gx(1-x) T_{out}'$.
Using an integrating factor,
\begin{equation}\label{eq28}
\left( Q(x) \left( \frac{x}{1-x} \right)^{r_0/g} e^{-\frac{\nu/g}{1-x}} \right)' = -\left( \frac{x^{-1+r_0/g}}{(1-x)^{1+r_0/g}} \right) e^{-\frac{\nu/g}{1-x}}.
\end{equation}
we obtained,
\begin{equation}\label{eq29}
T_{out}'(x)=  \frac{(1-x)^{-1+r_0/g}}{gx^{1+r_0/g}} e^{\frac{\nu/g}{1-x}} \int_x^1 dt \ \frac{t^{-1+r_0/g}}{(1-t)^{1+r_0/g}} e^{-\frac{\nu/g}{1-t}}.
\end{equation}
This solution satisfies the required outer  boundary condition $T_{out}'(1)=1/\nu$.
Defining $q \equiv (\nu t)/g(1-t)$ one may solve the integral,
\begin{equation}\label{eq30}
\int_x^1 dt \ \frac{t^{-1+r_0/g}}{(1-t)^{1+r_0/g}} e^{-\frac{\nu/g}{1-t}} = \left(\frac{g}{\nu}\right)^{r_0/g} e^{-\nu/g} \int_{\frac{\nu x}{g(1-x)}}^{\infty} dq \ e^{-q} q^{-1+r_0/g} =  \left(\frac{g}{\nu}\right)^{r_0/g} e^{-\nu/g} \Gamma \left( r_0/g,\frac{\nu x}{g(1-x)} \right),
\end{equation}
where $\Gamma(a,x)$ is the incomplete Gamma function. Accordingly,
\begin{equation}\label{eq31}
T_{out}'(x)=  \frac{(1-x)^{-1+r_0/g}}{gx^{1+r_0/g}} e^{\frac{x \nu}{ g(1-x)}} \left(\frac{g}{\nu}\right)^{r_0/g} \Gamma \left( r_0/g,\frac{\nu x}{g(1-x)} \right).
\end{equation}
Large argument asymptotics of the incomplete Gamma function shows that $T'(x) \approx 1/(\nu x^2)$ as $x \to 1$, as expected.

To find $c_1$ we matched the two solutions, $T_{in}'$ and $T_{out}'$, when $1/G \ll x \ll 1$.   In the limit $Gx \gg 1$ Eq. (\ref{eq26b}) yields,
\begin{equation} \label{eq19a}
T_{in} (Gx \gg 1) \sim \frac{c_1}{r_0 N}\left( 1-\frac{1}{(Gx)^{r_0/g}}\right)+ \frac{g}{r_0^2}\left(1-\frac{1}{x^{r_0/g}} \right) - \frac{\ln x}{r_0} -\frac{N}{(Gx)^{r_0/g}}I(1),
\end{equation}
and,
\begin{equation}\label{eq19b}
T_{in}'(Gx \gg 1) \sim \left( \frac{c_1}{G^{r_0/g+1}}+\frac{1}{r_0}+\frac{r_0 I(1)}{g^2 G^{r_0/g-1}} \right) \frac{1}{x^{r_0/g+1}} - \frac{1}{r_0 x},
\end{equation}
with the  $I(x)$ defined above (\ref{Idef}).

Comparing Eq. (\ref{eq19b}) with the small $x$ asymptotic of (\ref{eq29}) one finds,
\begin{equation} \label{eq32a}
c_1 = G^{r_0/g+1} \frac{\Gamma(r_0/g)}{g} \left( \frac{g}{\nu} \right)^{r_0/g} -  \frac{G^{r_0/g+1}}{r_0}  -  \frac{G^2 r_0 I(1)}{g^2}.
\end{equation}
Note that $I(1)$ is a hypergeometric function, one can take $gN$ to infinity and approximate $$I(1) \approx -\frac{g^2}{r_0^2}(G)^{r_0/g-1}+\frac{\gamma_E+\ln G + \Psi(-r_0/g)}{N r_0},$$ where $\Psi$ is the diagamma function.

\subsection{Time to extinction for a single invader/mutant}

Here we assume that $n(t=0) = 1$, or $x(t=0) = 1/N$. This corresponds to the case where a homogenous $B$ population is invaded by a single $A$, or when (due to extremely rare mutations that never happens again) $B$ mutates into $A$.

 Since $I(1)-I(x)$ appears as the difference between two Beta functions, by taking  first $x \to 0$ and than $G \to \infty$,
\begin{equation}
I(1)-I(x) \approx \frac{g}{Nr_0^2}\left((G x)^{r_0/g}+r_0x^{r_0/g}\left[\gamma_E+N r_0 x + \ln G x +\Psi(-r_0/g)\right] \right) - \frac{(1+G x)^{r_0/g} \ln x}{N r_0}.
\end{equation}

Plugging (\ref{eq32a}) into (\ref{eq26b}), the mean lifetime of a single mutant $x = 1/N$ turns out to be
\begin{equation} \label{single}
T(1/N) =  \left( \frac{\Gamma(r_0/g)G^{r_0/g}}{r_0}\left( \frac{g}{\nu} \right)^{r_0/g} -\frac{\gamma_E+ln(g)+\psi(-r_0/g)}{r_0}\right)  \left( 1-(1+g)^{-r_0/g} \right) + \left(1+g\right)^{-r_0/g}.
\end{equation}

When $r_0>0$ the mean  persistence time is dominated by the first term of (\ref{single}) and increases like $(Ng)^{r_0/g}$. This behavior is demonstrated in panel (A) of Fig. \ref{continue}. When $r_0<0$ the time to extinction becomes $N$-independent in the large-$N$ limit.

\subsection{Time to extinction for macroscopic populations}

\begin{figure}
	\includegraphics[width=5cm]{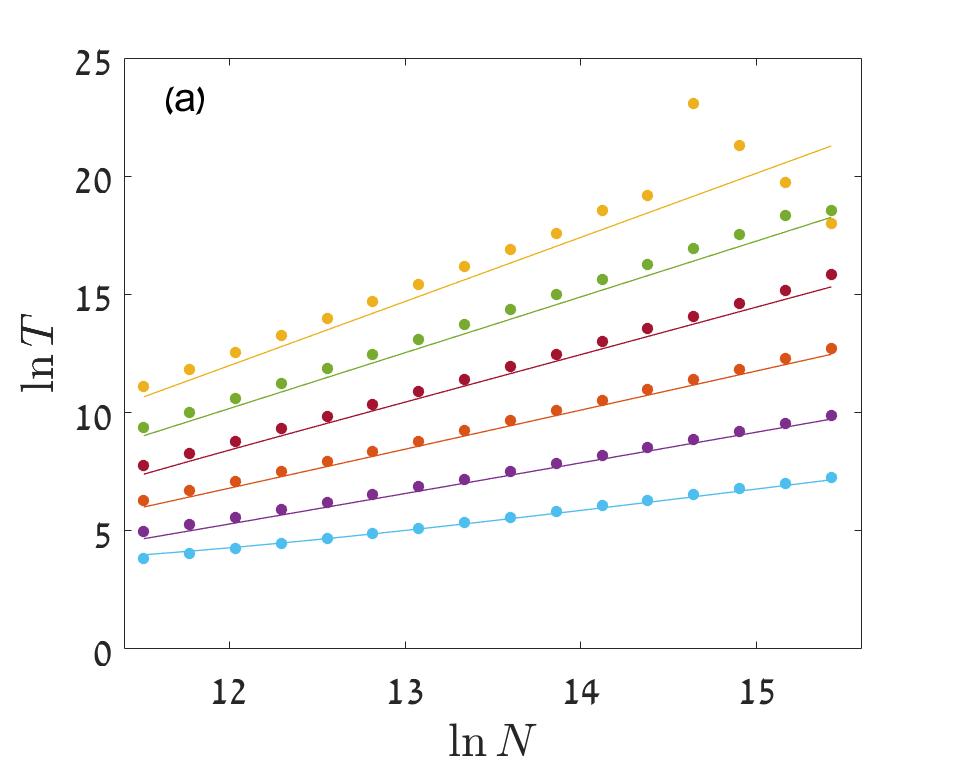}
	\includegraphics[width=5cm]{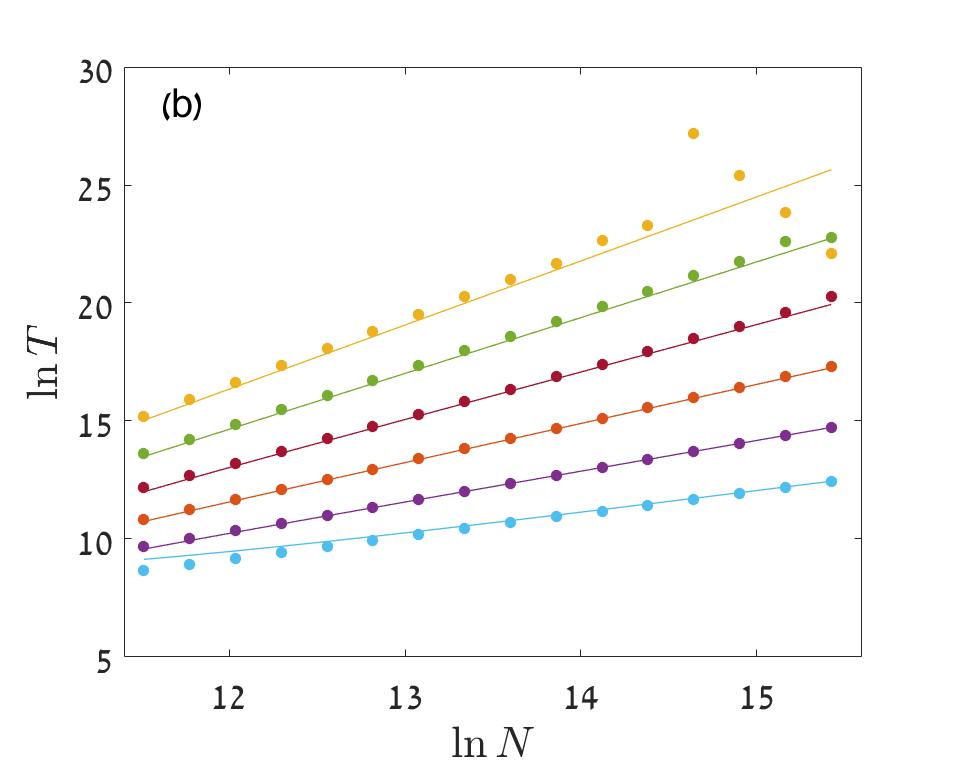}
	\includegraphics[width=5cm]{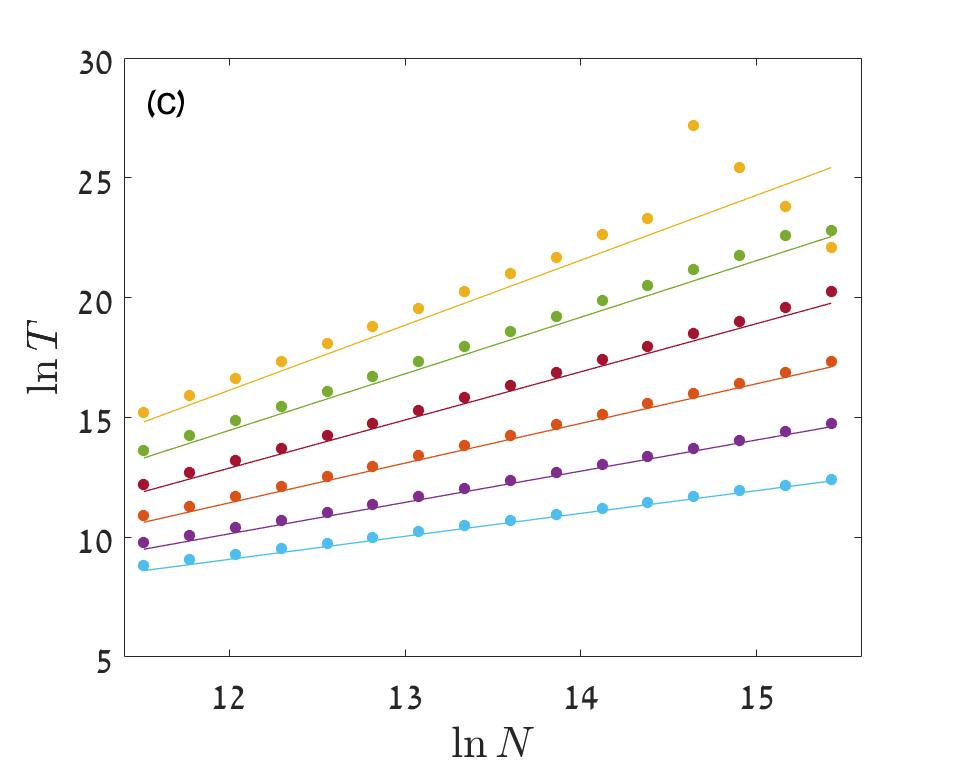}
	 \caption{The logarithm of $T$ vs. $\ln N$ for $x(t=0) = 1/N$ (panel A), $x(t=0)= 0.01$ (panel B) and $x(t=0) = 1$ (panel C). Dots were obtained from  numerical  solution of the exact BKE (\ref{eqb1}), as explained in the methods section.   The full lines are the prediction  of Eq. \ref{single} (panel A), Eq. \ref{large} (panel B) and Eq. \ref{full} (panel C). Parameters are $\tau =2$, $\tilde{\sigma} = 0.08$, $s_0 = 0.1$. Different colors represent different mutation probabilities:  $\nu = 0.088$ (light blue), $0.086$ (purple), $0.084$ (light red), $0.082$ (red), $0.08$ (green) and $0.078$ (yellow). Once the population becomes macroscopic its time to extinction depends only weakly on its size, so the differences between panels (B) and (C) are only minor.  At small $\nu$ and large $N$ (yellow, upper right) our numeric becomes less accurate.    } \label{continue}
\end{figure}

\begin{figure}
	\includegraphics[width=5cm]{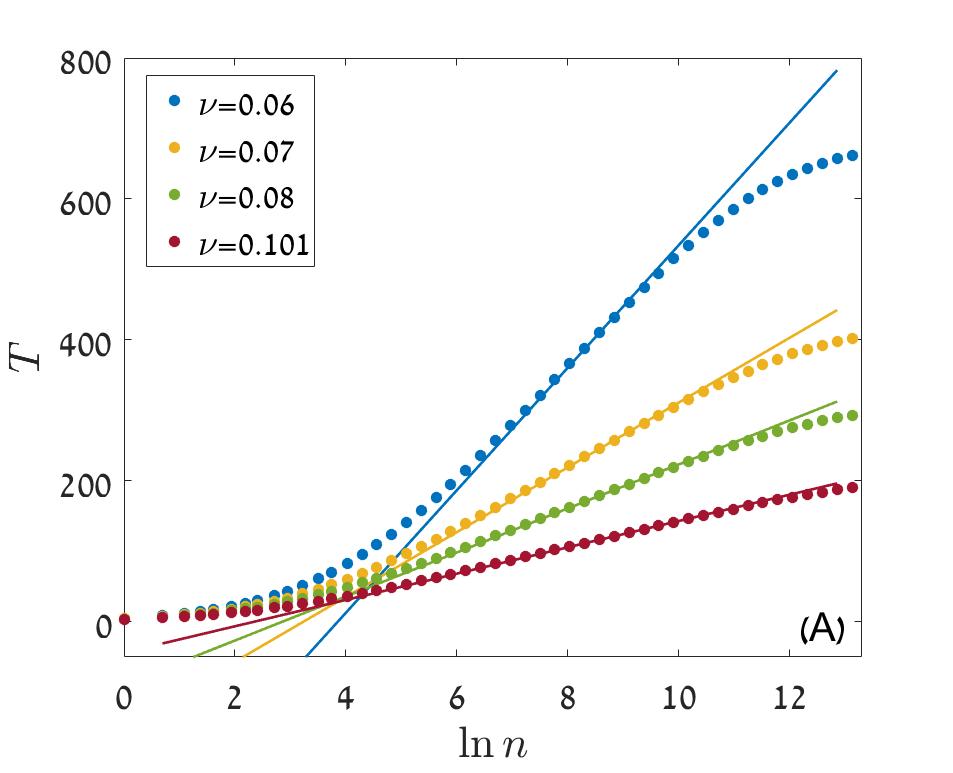}
	\includegraphics[width=5cm]{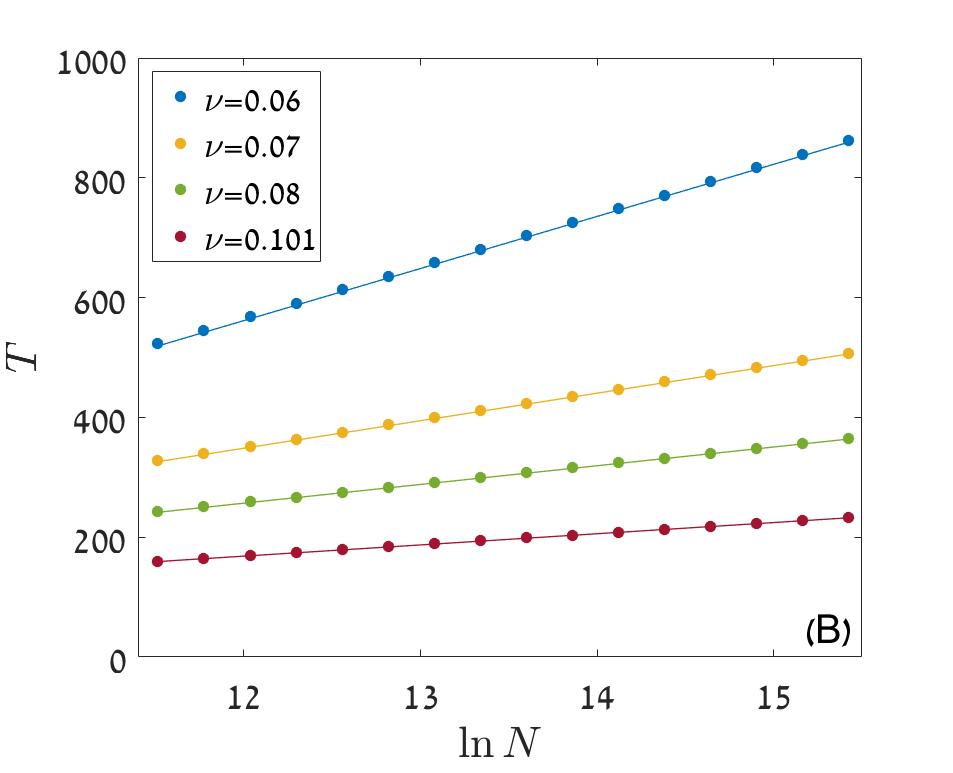}
	\caption{In the left panel, the time to extinction is plotted against the initial log-abundance when  $r_0<0$, i.e., in the inactive (logarithmic) phase. Parameters are $\tilde{\sigma} = 0.03$,  $s_0 = 0.05$, $N = 500000$ and $\tau = 2$. As in Fig. \ref{continue}, points are the numerical solution of the exact BKE while lines are analytic predictions. The lines represent the two last terms of Eq. (\ref{large}). The theory fits the numerical experiment in the bulk, where the asymptotic matching works.  Panel B (right) provides the results for $x(t=0) = 1$. The lines represent the predictions of Eq. (\ref{full}) and the agreement is very good. } \label{continuelog}
\end{figure}

For population of abundance $n$, if $ng \gg 1$  (or equivalently $Gx \gg 1$) the effect of demographic noise is weak. This condition allows one to approximate $I(1)-I(x)$ by neglecting $1$ with respect to $Gx$. Plugging the outcome into the equation in the matching regime $x \ll 1$ one gets,
\begin{equation} \label{large}
T_{in} = \left(G^{r_0/g}-x^{-r_0/g} \right) \frac{\Gamma(r_0/g)}{r_0} \left( \frac{g}{\nu} \right)^{r_0/g} - \frac{\ln Gx}{r_0} -\frac{\gamma_E-g/r_0+\Psi(-r_0/g)}{r_0}.
\end{equation}
Accordingly, the time to extinction grows, again, like $N^{r_0/g}$ if $r_0$ is positive~\cite{lande2003stochastic,kamenev2008colored}, as demonstrated in panel (B)  of Figure \ref{continue}. $T$  is logarithmic in $n=Nx$ when $r_0$ is negative as depicted in panel (A) of Fig \ref{continuelog}.

When $x$ becomes larger, one would like to write down the time to extinction as
\begin{equation}
T(x) = \int_0^\xi T'_{in}(x) \ dx + \int_\xi^x T'_{out}(x) \ dx,
\end{equation}
where $\xi$ is somewhere in the overlap region $1/G \ll \xi \ll 1$. Since $T'_{out}(x)$ is $N$-independent, the only terms that diverge with $N$ come from the first integral.

Note that, for $r_0>0$  the large $N$ asymptotic of the mean lifetime  of a single mutant (\ref{single}) differs from the corresponding quantity for large population (\ref{large}) only by the factor $1-(1+g)^{-r_0/g}$, which converges to $r_0$ when $g \ll 1$. This factor represents the chance of establishment for a single beneficial mutant/immigrant.  The probability of absorption at zero, for a random walker that moves to the right with probability $1/2 + r_0/4$ and to the left with probability $1/2 -r_0/4$, is given by $1-r_0$ for $r_0 \ll 1$, so $r_0$ is the chance to escape the absorbing state. The time to extinction is $N$ dependent only when the upper bound  (the carrying capacity $n^*$, which is linear in $N$) determines the chance of extinction. For a single mutant the chance that the extinction happens because of this upper bound, and not as a result of the random walk in the log-abundance space, is equal to the chance of establishment  $1-(1+g)^{-r_0/g} \approx r_0$.

\subsection{Time to extinction from maximal capacity}

To calculate $T(1)$ one can compare $T(x) = T(1)+\int_{1}^{x} T'_{out}(t) dt$ in the matching regime ($1/G \ll x \ll 1$) with Eq. (\ref{large}). Defining $q \equiv (\nu t)/g(1-t)$ we solve the integral and approximate the solution for $x \ll 1$
\begin{equation}
\frac{1}{g}\int_{\infty}^{(\nu x)/g(1-x)} e^q q^{-(1+r_0/g)}\Gamma(r_0/g,q) dq \approx \frac{\Psi(r_0/g)- \ln (\nu x/g)}{r_0} - \frac{\Gamma(r_0/g)}{r_0}\left(\frac{\nu x}{g}\right)^{r_0/g}.
\end{equation}
a comparison between this result and  Eq. (\ref{large}) yields,
\begin{equation}\label{full}
T(1) = \frac{\Gamma(r_0/g)}{r_0}\left(\frac{g}{r}G\right)^{r_0/g} +\frac{2\Psi(-r_0/g)+\ln (\nu/Gg)-\gamma_E+g/r_0}{r_0}.
\end{equation}
This result are demonstrated in panel (C) of Fig. \ref{continue} (for $r_0>0$) and in panel (B) of Fig. \ref{continuelog} for $r_0<0$.

\section{Lifetime at criticality} \label{critical}

Through the last  section we have assumed that in the power-law phase the term $G^{r_0/g}$ dominates, so one may neglect terms which are logarithmic in $N$, or ${\cal O}(1)$. This is the case for each $r_0>0$ as long as $N>n_c$ \cite{cvijovic2015fate} where
\begin{equation}
n_c = \frac{e^{g/r_0}-1}{g}.
\end{equation}
$n_c$ marks the point above which the deterministic grows, associated with $r_0$, dominates both demographic and environmental fluctuations. As $r_0$ decreases while $g$ is kept fixed, $n_c$ grows exponentially and, for any fixed $N$, the system enters the regime where the deterministic growth term becomes negligible. This is the critical regime that separates the power-law and the $N$-independent phase, and here we discuss the mean time to extinction in this regime.

When the linear growth rate vanishes ($r_0=0, \ s_{eff} = -\nu$), Eq. (\ref{eq9}) takes the form,
\begin{equation}\label{eq10}
\left(- \frac{\nu x}{1-x} + g (1-2x) \right)T'(x) + \left(\frac{1}{N} + g x(1-x) \right) T''(x)  = -\frac{1}{x(1-x)}.
\end{equation}

In the outer regime the $1/N$ term is negligible and,
 \begin{equation} \label{14}
\left(- \frac{\nu x}{1-x} + g (1-2x) \right)T'(x) + g x(1-x) T''(x)  = -\frac{1}{x(1-x)}.
\end{equation}
The definition $Q = g x(1-x)T'_{out}$ allows one to write a first order equation for $Q$,
\begin{equation} \label{15}
Q'- \frac{\nu Q}{g (1-x)^2}   = -\frac{1}{x(1-x)},
\end{equation}
and this equation may be solved for $T'$,
\begin{equation} \label{out1}
T_{out}'(x) = c_2 \frac{e^{\nu/g(1-x)}}{x(1-x)} + \frac{e^{\nu x/g(1-x)}}{gx(1-x)}Ei\left(\frac{\nu x}{g(1-x)}\right).
\end{equation}
To satisfy the right boundary condition $T'(1)=1/\nu$, $c_2$ must vanish.

In the inner regime where $1-x \approx 1$ we can use the definition  $G \equiv gN$ and $W = (1+G x)T_{in}'$ to write,
 \begin{eqnarray} \label{12}
 W' - \frac{N \nu x}{1+G x} W &=& -\frac{N}{x}  \nonumber \\
 \left(We^{-\nu x/g}(1+Gx)^{\nu /g^2N}\right)'&=&-N\frac{e^{-\nu x/g}(1+Gx)^{\nu /g^2N}}{x}
 \end{eqnarray}
As long as $\nu/g^2 \ll N/\ln N$ (that is, $g$ is not vanishingly small), the term $(1+Gx)^{\nu/Ng^2}$ tends to one when $N$ is large. Therefore,
\begin{equation}
T'_{in} = \frac{ Ne^{\nu x/g}}{1+G x} [ Ei(\nu x/g)-c_1 ],
\end{equation}
and the matching between $T'_{in} (Gx \gg 1)$  and $T'_{out}(x \ll 1)$ dictates $c_1 = 0$.

Another integration, plus the boundary condition $T(x=0) = 0$, yields,
\begin{equation} \label{18}
T_{in} \approx  -\frac{1}{g} \left[\gamma_E + \log \left(\frac{\nu x}{g} \right) \right] \log(1+G x) -\frac{ Li_2(-Gx)}{g},
\end{equation}
where the dilogarithmic function $Li_2(-x) \sim \ln^2(x)$ when $x$ approaches infinity. Accordingly, for a single muntant ($x = 1/N$) the only $N$ dependence comes from the $\ln(\nu x /g)$ term and the time to extinction is logarithmic in $N$,
\begin{equation} \label{17}
T(1/N) \approx \frac{\ln(1+g)}{g} \ln N.
\end{equation}
On the other hand, when $Gx \gg 1$ the $\ln^2 N$ behavior dominates~\cite{vazquez2011temporal}. For example, when  $x \sim N^{-\beta}$ with  $\beta < 1$, the argument of the dilogarithmic function approaches minus infinity with $N$ and  the leading contribution to the lifetime is,
\begin{equation} \label{17a}
T(x \sim N^{-\beta}) \sim \frac{(1-\beta^2)}{2g} \ln^2 N
\end{equation}

\begin{figure}
	\centering{
		\includegraphics[width=5cm]{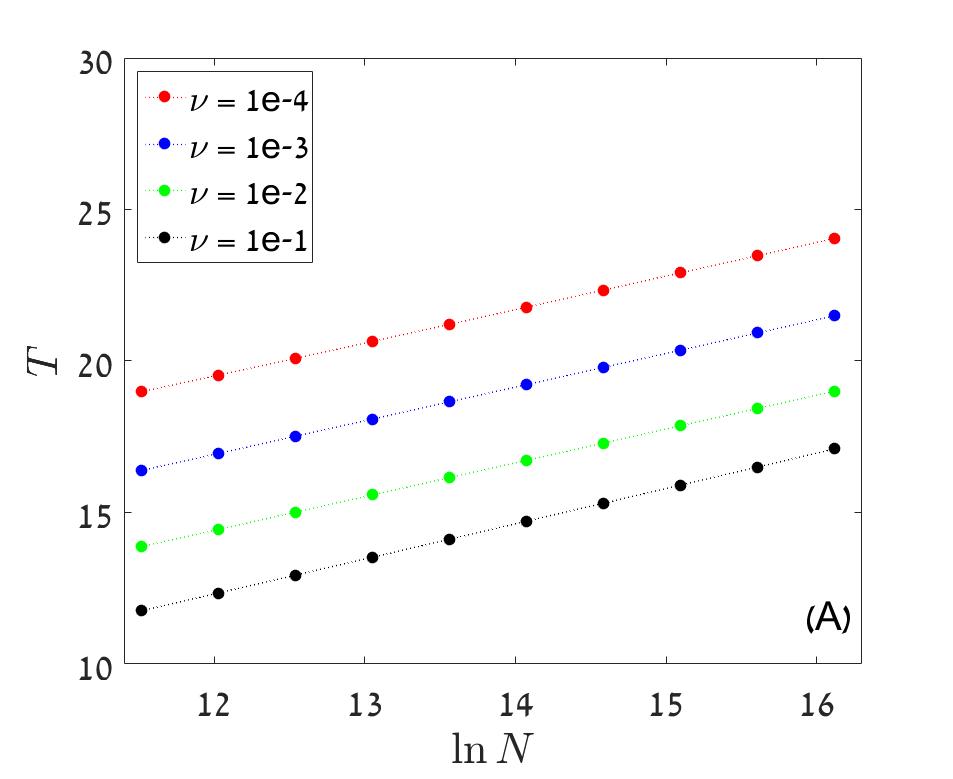}
		\includegraphics[width=5cm]{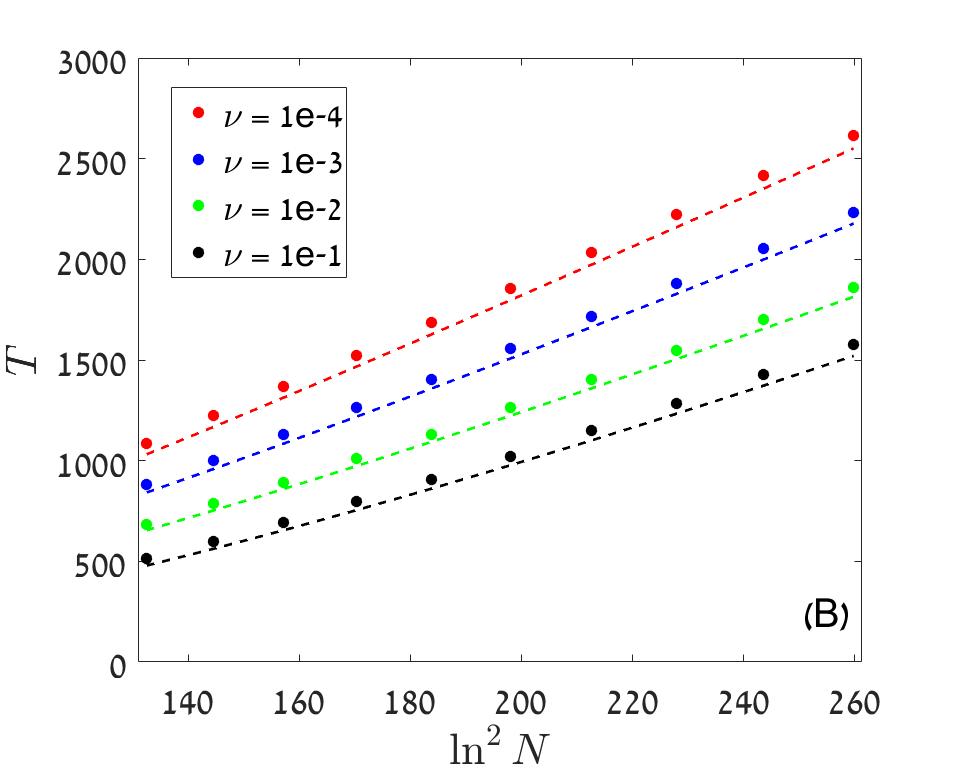}
		\includegraphics[width=5cm]{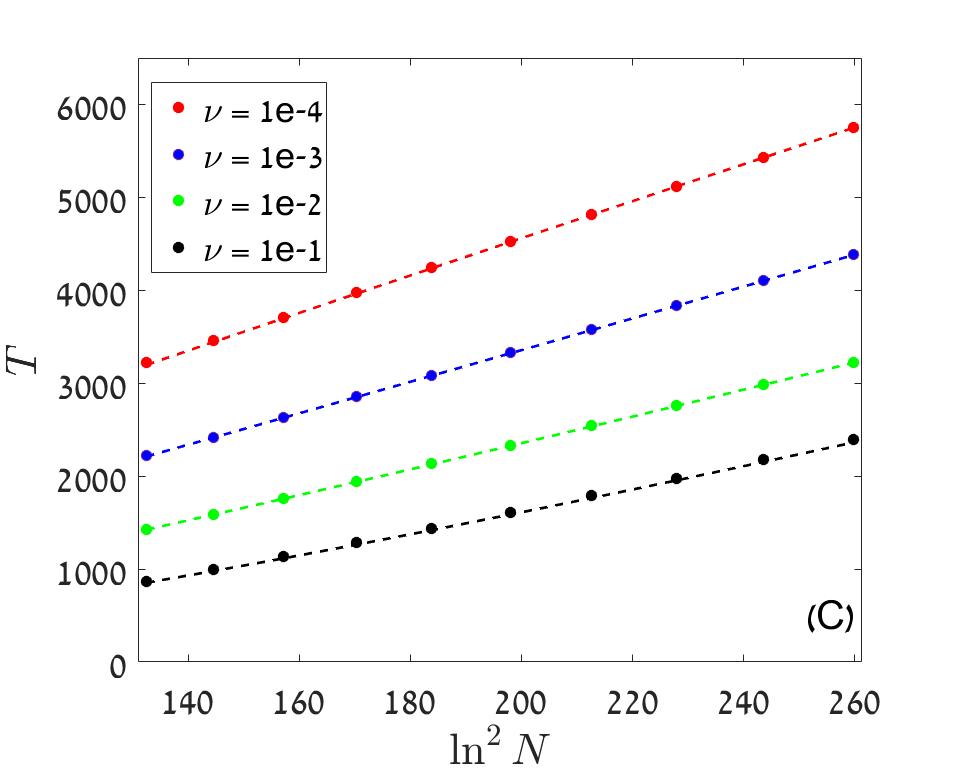}}
	\caption{The time to extinction as a function of $N$ at the transition between the inactive (logarithmic) phase and the power-law region, $r_0=0$, for finite environmental stochasticity $\tilde{\sigma} = 0.21$ and $\tau = 1.25$.  Different colors correspond to different $\nu$s, see legend. Points were obtained from a numerical solution of Eqs. (\ref{eqb1}), dashed lines are analytic solutions. Panel A (left) provides the results for a single mutant ($n(t=0) = 1$). Here Eq. (\ref{17}) does not give the exact numbers because of discretization effects, still the linear dependence on $\ln N$ is observed, as demonstrated by the dotted straight lines, plotted only to guide the eye. In panel B (middle),  $n(t=0) = \sqrt{N}$ and the dashed lines represent the predictions of Eq. (\ref{18}),  the agreement is very good. In panel C (right) the results for maximal capacity  $n(t=0) = N$ are compared with  the predictions of Eq. (\ref{fullcc}) (dashed) and the agreement is perfect}\label{fig2}
\end{figure}

As before, since $T_{out}'$ is $N$-independent, the same $\ln^2 N$ behavior characterizes the outer regime.  The validity of these formulas is demonstrated in Figure \ref{fig2}.

 $T(x)$ may be calculated [up to a constant, which is $T(1)$] in a different way, by integrating  $T'_{out}$ from 1 to $x$.  With the definition $q \equiv (\nu t)/g(1-t)$  one  gets, when $x \ll 1$
$$\frac{1}{g}\int_{\infty}^{(\nu x)/g(1-x)} \frac{e^q}{gq}Ei(q) dq \approx -\frac{1}{4g}\left(2\gamma_E^2+\pi^2+4 \gamma_E \ln \nu/g +2 \ln^2 \nu/g -4\left[\gamma_E+\ln \nu/g \right]\ln x + 2 \ln^2 x\right).$$
Subtracting this from  Eq. (\ref{18}) one gets  the time to extinction at maximal capacity,
\begin{equation} \label{fullcc}
T(1) = \frac{1}{4g} \left(2\gamma_E+5\pi^2/2+4\gamma_E\ln \nu/g+2\ln^2 \nu/g-2\left[2\gamma_E+2\ln \nu/g -\ln G \right]\ln G \right).
\end{equation}
See Fig. \ref{fig2}, panel (C).

\section{Critical system with pure demographic noise} \label{hom}

To provide a comprehensive outlook, we would like in this section to calculate the mean time to extinction at the transition point for a system under pure demographic noise,  $r_0=g=0$.

Eq. (\ref{eq9}) now takes the form,

 \begin{equation} \label{7}
\frac{1}{N}T'' - \frac{\nu x}{1-x} T' = -\frac{1}{x(1-x)}.
\end{equation}

\begin{figure}
	\centering{
		\includegraphics[width=5cm]{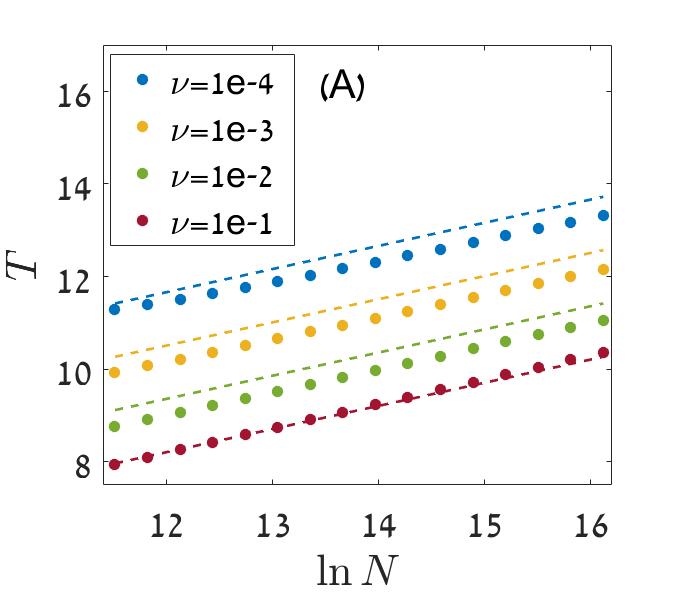}
		\includegraphics[width=5cm]{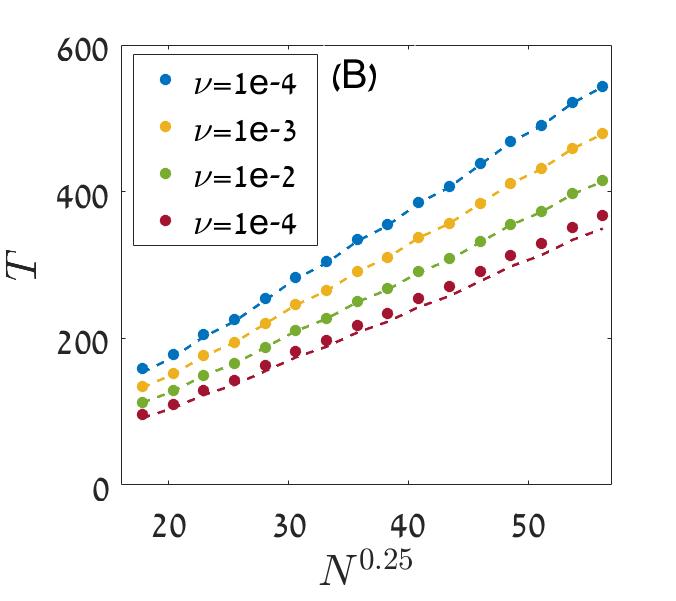}
		\includegraphics[width=5cm]{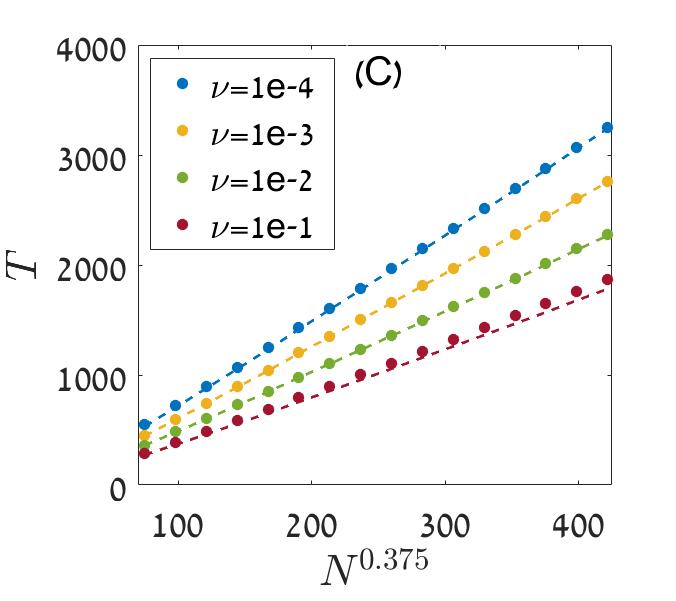}
		\includegraphics[width=5cm]{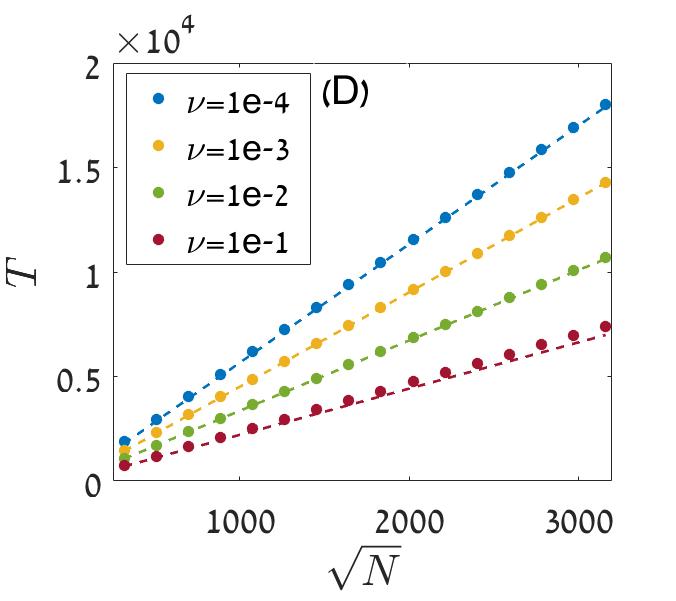}
		\includegraphics[width=5cm]{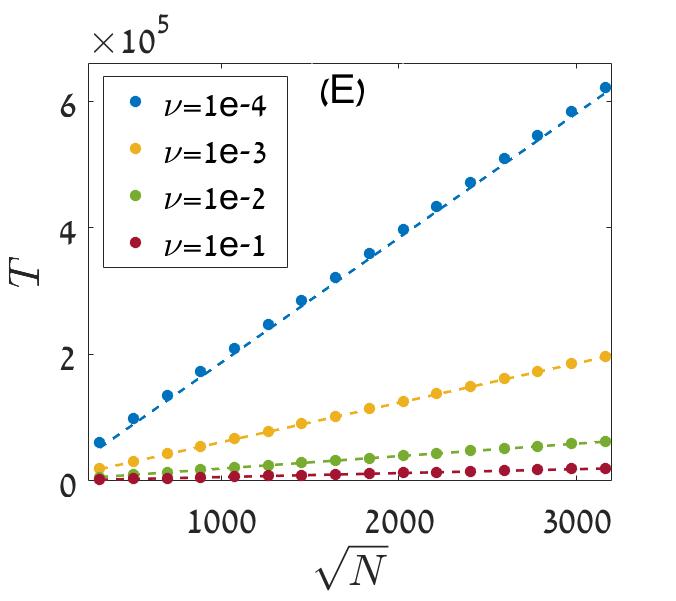}}	
	\caption{Time of extinction as function of $N$ where $r_0=0$  without environment stochasticity ($\tilde{\sigma}=0$).  Points were obtained from a numerical solution of Eqs. (\ref{eqb1}). Panel A provides the results for a single mutant ($n(t=0) = 1$), the dashed lines represent the corresponding predictions of Eq. (\ref{ccmut}). Panels B-D provides the results for $n(t=0) = N^{-\beta}$ where  $\beta=0.75$ (panel B), $\beta=0.625$ (panel C) and $\beta=0.5$ (panel D). the dashed lines represent the corresponding predictions of Eq. (\ref{rescrit0}) with $C_1 = 1-(\gamma_E+\ln \nu/2)/2$ and $C_2 =1/2$. Panel E provides the results for maximal capacity ($n(t=0) = 1$), the dashed lines represent the corresponding predictions of Eq. (\ref{11}). The constants $C_1$ and $C_2$ calculated by approximate $T'_{in}$ (Eq. \ref{8}) for $x^2 \ll 1/N\nu$.}\label{figcc}
\end{figure}

In the inner regime where $1-x \approx 1$ we can write,
 \begin{eqnarray} \label{8}
T_{in}'' - N \nu x T_{in}' = -\frac{N}{x} \quad T_{in}(0) = 0 \nonumber \\
\left(T_{in}'e^{- N \nu x^2/2}\right)'=-N\frac{e^{- N \nu x^2/2}}{x}  \\
T_{in} = c_1 \int_0^{x} e^{N \nu t^2/2 } \ dt + \sqrt{\frac{N}{8 \nu}} \int_0^{N \nu x^2/2} \frac{e^z E_i(z)}{\sqrt{z}} \ dz. \nonumber
\end{eqnarray}
In the outer regime $N \nu x \gg 1$, $T''$ is negligible and both the remaining equation and the  boundary condition at $x=1$ are satisfied by,
\begin{equation} \label{9}
T_{out}' = \frac{1}{\nu x^2}.
\end{equation}
Accordingly,
\begin{equation} \label{10}
T_{out} = c_2 -\frac{1}{\nu x}.
\end{equation}
To match $T_{in}$ with $T_{out}$ $c_1$ has to vanish since the first integral in the last line of (\ref{8}) diverges in the limit $N \nu x \to \infty$. Evaluating the large $N \nu x^2/2$ limit of the other integral one finds that $$c_2 = \sqrt{\frac{N \pi^3}{8 \nu}}, $$ so the mean time to extinction, starting at $x=1$, is
\begin{equation} \label{11}
T_{out}(x=1) = \sqrt{\frac{N \pi^3}{8\nu}} - \frac{1}{\nu}.
\end{equation}
This agrees with the result obtained by \cite{doering2005extinction} and with the numerics (panel (E) of Fig. \ref{figcc}).

On the other hand, the time to extinction for a single mutant is obtained by plugging $x = 1/N$ in $T_{in}$,
\begin{equation}\label{ccmut}
T(1/N) = 1 - \frac{\gamma_E+\ln \left(\nu/2\right)-\ln(N)}{2}.
\end{equation}
Panel (A) of Fig. \ref{figcc} demonstrated the accuracy of this approximation.

 In general, when  $x \sim N^{-\beta}$ where $1/2 \le \beta \le 1$, the time to extinction scales like
 \begin{equation} \label{rescrit0}
 T \sim N^{1-\beta} [C_1+C_2 \ln(N^{2\beta -1})],
  \end{equation}
  where $C_1$ and $C_2$ are some constants. The fit of this formula to numerical results (panels B-D of Fig. \ref{figcc}) is very good.

\section{Failure of the continuum approximation and a WKB approach} \label{WKBsec}

The analysis so far suggests that when $r_0>0$ the time to extinction grows like $N^{r_0/g}$. This cannot be a general statement. In systems with pure demographic stochasticity it is already known that $T$ grows exponentially with $N$ when $r_0>0$ and $N$ is taken to be large~\cite{lande2003stochastic,kessler2007extinction,ovaskainen2010stochastic}. If $\sigma<r_0$ the system jumps between two states, both with positive growth rate, so even in the worst case scenario, when it is stacked for a long time in the negative $\sigma$ state, the time to extinction is still exponential in $N$. Accordingly, the power law behavior must cross over to an exponential behavior when $ r_0 \to \sigma$.

This simple argument points out to  the collapse of the continuum approximation presented above. When the continuum approximation holds, the characteristics of the system depend on $g = \sigma^2 \tau/2$, which is the effective diffusion constant in  log-abundance space. Conversely, when the system reaches the point where $\sigma=r_0$ its qualitative behavior changes dramatically, since now the linear growth rate is always positive and extinction happens only due to demographic noise. This feature is independent of the value of $\tau$, meaning that the diffusion approximation must break down somewhere inside the power-law region.

To analyze the system when the continuum approximation fails, we adopt a version of the WKB analysis presented and discussed in \cite{kessler2007extinction,meyer2018noise}. We shall neglect the demographic noise and replace it (as in \cite{lande2003stochastic,hidalgo2017species}) by an absorbing boundary condition at $x=1/N$.

We begin with our logistic equation,
\begin{equation} \label{detwkb}
\dot{x} = (r_0 \pm \sigma) x - (s_{eff} \pm \sigma) x^2,
\end{equation}
and for simplicity we would like to slightly modify the process that governs the environmental dynamics. Until now the environment persistence time was taken from an exponential distribution with mean $\tau$. From now on we consider an environment that stays in the same state (plus or minus $\sigma$) for $\tilde{\tau}$ generations and than  \emph{may } switch, with probability $1/2$, to the other state (minus or plus $\sigma$), while with probability $1/2$ it stays in the same state. Accordingly, the dynamics is still random but now the persistence time is picked from a geometric distribution with mean $\tilde{\tau}$; below we will explain the relationships between $\tilde{\tau}$ and  $\tau$.

 From Eq. (\ref{detwkb}) one finds that, if the system reaches $x$ at certain time $t$, then one time increment before, i.e., at  $t- \tilde{\tau}$, it was either at $\ell_+(x)$ or $\ell_{-}(x)$ where,
\begin{eqnarray} \label{ell}
\ell_{+} = \frac{x^*_+}{\left (\frac{x^*_+}{x}-1 \right) e^{\tilde{\tau}  (r_0+\sigma)}+1} \\ \nonumber \ell_{-} = \frac{x^*_-}{\left (\frac{x^*_-}{x}-1 \right) e^{\tilde{\tau}  (r_0-\sigma)}+1},
\end{eqnarray}
and the quantity $$x^*_\pm = 1-\nu/(s_{eff} \pm \sigma),$$ is the nonzero fixed points of the plus and the minus state, correspondingly.  When $r_0 < \sigma$ the value of $x_{-}$ is non physical, either below zero or above one, but $\ell_{-}$ is between zero and one.

\begin{figure}[h]
\includegraphics[width=10cm]{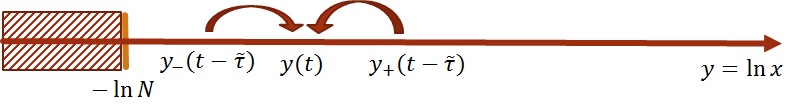}
\caption{ An illustration of of the dynamics considered using the WKB approximation. The dynamics takes place on the $y=\ln x$ axis, and the chance of the system to be at $y$ when the time is $t$  is completely determined by its chance to be at $y_{\pm}$ at $t-\tilde{tau}$. The values of $y_\pm$ are calculated from the deterministic equation \ref{detwkb}, without taking demographic stochasticity into account, so there are only two initial conditions that map to $y(t)$ - each correspond to one state of the environment. Without demographic noise, this dynamics yields a stable probability distribution function $P(y)$ if $r_0>0$. The only effect of demographic noise that we take into account is the possibility of extinction, which is proportional to the support of $P(y)$ on the extinction zone $x<1/N$, or $y<-\ln N$.  }\label{WKBillustration}
\end{figure}

 Now let us define $y \equiv \ln x$ and $y_\pm \equiv \ln \ell_\pm$.  The probability to find the system at the log-density $y$ at time $t$, $P(y,t)$ satisfies the master equation,
\begin{equation} \label{master}
\frac{dP(y,t)}{dt} = \frac{1}{2} \left[-2P(y) + P(y_+) + P(y_-) \right].
\end{equation}

As in \cite{kessler2007extinction}, we assume the existence of a quasi stationary probability distribution function $P(x)$. Although the system "leaks" to extinction from the state with one individual, we consider this leak to be extremely weak,  so the system equilibrates to its quasi steady state on timescales that are much shorter than the mean time to extinction (see Figure \ref{WKBillustration}). This allows one to solve Eq. (\ref{master}) neglecting  $dP/dt \approx 0$. The rate of extinction is then estimated from  the probability to find the system with one individual, i.e., to be in the region  $0<x<1/N$, or $-\infty<y<-\ln N$,
$$ {\rm Rate} \sim \int_{-\infty}^{-\ln N} P(y) \ dy,  $$
and the mean time to extinction is the inverse of this rate. The large-$N$ asymptotics of the extinction rate depends only on the small-$x$ asymptotics of the quasi-stationary probability function, the behavior of $P(y)$ at larger $y$-s determines only the normalization factor, but this factor is independent of $N$.

In the extinction regime $x$ is vanishingly small and $\ell_\pm \approx x e^{-\tilde{\tau}(r_0\pm \sigma)}$. Accordingly, for $x \ll 1$  the quasi-stationary  state satisfies,
\begin{equation}
P(y-\tilde{\tau}[r_0+\sigma]) + P(y-\tilde{\tau}[r_0-\sigma])=2 P(y).
\end{equation}

Now we implement a WKB approach.  Instead of expanding $P(y_\pm)$ to second order in $\Delta y$ (this will give us the continuum Fokker-Planck equation and the power-law behavior of the continuum limit) we expand the logarithm of $P$ in small $\Delta y$. The breakdown of the continuum approximation suggests that $P_n$ varies significantly over the integers, so the approximation $P(x+1/N) \approx P(x) +P'/N +P''/(2N^2)$ fails. Still, the logarithm of $P$ may be a smooth enough function.

Accordingly, we write $P(y) = e^{S(y)}$ and implement the continuum approximation to $S$, replacing $S(y+\Delta y)$ by $S(x)+\Delta y S'(x)$,  so $S'(x)$ is obtained as  a solution of the transcendental equation
\begin{equation} \label{trans}
\exp \left(- \tilde{\tau}   r_0 S'\right) \cosh\left( \tilde{\tau} \sigma S' \right) = 1.
\end{equation}
This equation does not depend on y, so $S'=q$ where $q$ is some constant. Accordingly  $S \sim q y$, so $P \sim \exp(q y)$, the rate satisfies ${\rm Rate} \sim N^{-q}$  and the time to extinction behaves like $$T \sim N^q.$$

When $r_0 \ll \sigma$ one expects $q \ll 1$. When this is the case both $q \tilde{\tau} r_0$ and $q \tilde{\tau} \sigma$ are small numbers and  Eq. (\ref{trans}) yields,
\begin{equation}  \label{res1}
q = \frac{2 r_0}{(\sigma^2+r_0^2) \tilde{\tau} } \approx \frac{2 r_0}{\sigma^2 \tilde{\tau} },
\end{equation}
where the last approximation reflects a self consistency requirement for $q \tilde{\tau} r_0 \ll 1$. On the other hand if $q \tilde{\tau} \sigma$ is large,
\begin{equation} \label{res2}
q = \frac{\ln 2}{\tilde{\tau}(\sigma-r_0)}.
\end{equation}

The case (\ref{res1}) corresponds to the regime where the continuum approximation holds. In that case the typical history that takes the system to extinction is a random walk in the log-abundance space (see discussion in Section \ref{pdf} and in particular Figure \ref{figtraj}). To translate $\tilde{\tau}$ to the parameter $\tau$ with which the our model was defined in the former sections, one has to compare the variance of the sum of $M$  steps of length $\tilde{\tau}$ and random direction, with the variance of the sum of $M$ numbers picked independently from an exponential distribution with mean $\tau$, half of them with plus sign and half with a minus sign. This implies that $\tilde{\tau}=\tau$ and
\begin{equation} \label{diff}
T \sim N^{r_0/g},
\end{equation}
 as expected.

In the other extreme, Eq. (\ref{res2}), the random walk of the system along the log-abundance axis is strongly bias to the right (towards larger abundance state). In that case  extinction occurs due to a (rare) long  sequence of bad years. In Appendix \ref{path} we argue that the most probable length of such a path scales with $\ln N$.  When $N$ is large this implies that $\tilde{\tau}$ must be compared with the tail of the corresponding exponential distribution, in which case $\tilde{\tau} = \tau \ln 2$, hence
 \begin{equation} \label{nondiff}
 T \sim N^{\frac{1}{\tau(\sigma-r_0)}},
 \end{equation}
so the time diverges as $r_0 \to \sigma$, as suggested above.

\begin{figure}
	\includegraphics[width=8cm]{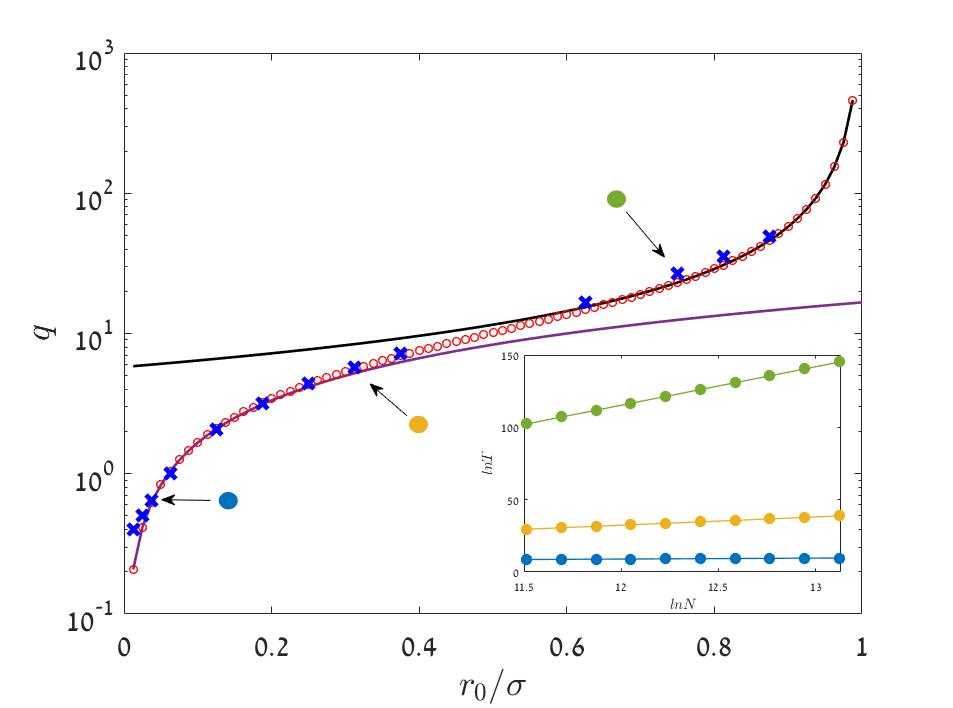}
	\caption{In the power-law $T \sim N^{q}$. The main panel shows  $q$ vs. $r_0/\sigma$ as obtained from numerical solution of Eq. (\ref{trans}) (red open circles), in comparison with the asymptotic expressions for the diffusive regime [Eq. (\ref{diff}), purple line] and in the large $r_0$ regime [Eq. (\ref{nondiff}), black line]. In the inset we present results for $T(N)$ as obtained from the numerical solution of the exact backward Kolomogorov equation for $r_0 = 0.003$ (blue circles)  $0.025$ (yellow) and $0.06$ (green). By fitting these numerical results (full lines) one obtains the actual power $q$, and the outcomes are represented by blue $X$s in the main panel (the $X$s that correspond to the three specific cases depicted in the inset are marked by arrows in the main panel). In general the predictions of the WKB fit quite nicely the numerical outcomes, and the slight deviations in the low $r_0$ region are due to the prefactors of the power law, and in these cases the numerical $T(N)$ graph fits perfectly the predictions of Eq. (\ref{large}). All the results here were obtained for  $\sigma = 0.08, \ \tau = 3/2, \ \nu = 0.04$.   }\label{fig4}
\end{figure}

Beside these limits, The transcendental  equation  (\ref{trans}) has to be solved numerically. In figure \ref{fig4} these numerical solutions are compared with the results obtained  from a numerical solution of the BKE and with the asymptotic expressions (\ref{diff}) and (\ref{nondiff}).

This WKB analysis covers the power-law regime where the linear growth rate may become negative and the time to extinction is related to the chance for a sequence of bad years. The procedure breaks down at $\sigma = r_0$. Above this point there are no bad years anymore and extinction happens only due to demographic stochasticity, as discussed in the next section.

\section{The power-law-exponential transition and the exponential phase} \label{expsec}

In the logarithmic and in the power law phase, environmental fluctuations may lead to extinction via a sequence of bad years. The minimal sequence of bad years that cause extinction is of order $\ln N$, and (see discussion in Appendix \ref{path}) this is also the typical path to extinction when $r_0$ is relatively large. When  the linear growth rate  is positive and $\sigma = 0$ (no environmental stochasticity) extinction is  still the ultimate fate of the system but now the dominant mechanism is  demographic noise, so the time to extinction grows exponentially with $N$, as discussed recently by many authors \cite{elgart2004rare,assaf2006spectral,ovaskainen2010stochastic}. Here we would like to present a brief qualitative discussion of the effect of environmental variations in this exponential phase, i.e., when $\sigma < r_0$ so the linear growth rate is always positive.

To begin, let us consider the transition points between the power-law and the exponential phase, i.e., the line $\sigma = r_0$ in Figure \ref{fig1}.   At the transition point the system fluctuates between two states, in one of them the linear growth is positive and the time to extinction is exponential in $N$, while in the other state the linear growth rate is zero.

\begin{figure}
	\includegraphics[width=8cm]{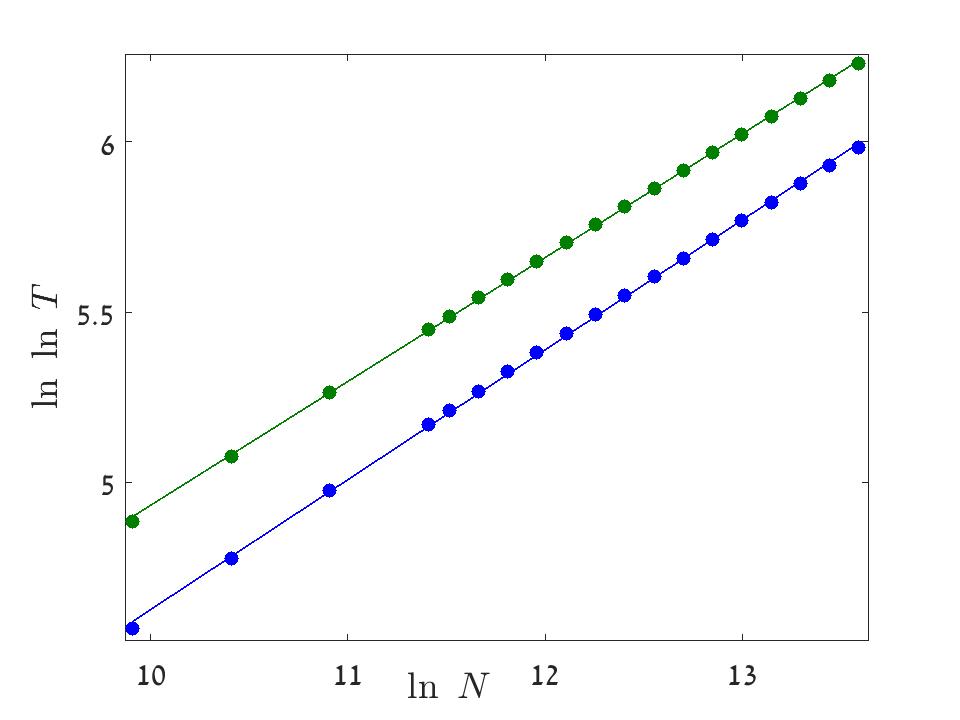}
	\caption{Stretched exponential relationships between the mean time to extinction $T$ and the carrying capacity $N$ at the power-exponential transition point $r_0=\sigma$.  $\ln \  \ln \ T$, as obtained from numerical solutions of the BKE (filled circles), is plotted against $\ln \ N$ and the linear fits (straight lines) suggest slopes around $\gamma \approx 0.37$. Parameters are $\tau =1$, $\tilde{\sigma} = 0.4$ and $\nu =0.2$ (green), $0.1$ (blue).     }\label{fig3}
\end{figure}

As we have shown in section \ref{hom}, when the environment does not fluctuate and the linear growth rate is zero (the transition point $r_0 = 0$, $\sigma =0$  at the origin of  Fig. \ref{fig1}) the time to extinction, when the initial fraction of the population is ${\cal O} (1)$, scales like the square root of $N$. Accordingly,  the most plausible route to extinction for a system that jumps between $r_0-\sigma=0$ and $r_0+\sigma >0$ is a sequence of   ${\cal O} (\sqrt{N})$ marginal years, an event that occurs with probability $\exp(-\sqrt{N}/\tau)$. The rate of  extinction in the positive growth state decays exponentially with $N$, so it is subdominant in the large $N$ limit. Given that, at the transition point one expects a stretched exponential behavior of the time to extinction
\begin{equation}
T \sim e^{\sqrt{N \pi^3/8\nu}}.
\end{equation}
As seen in Figure \ref{fig3}, the behavior is indeed stretched exponential  $T \sim \exp(N^\gamma)$ but  $\gamma \approx 0.37$. We believe that this deviation has to do with the full probability distribution function $f(T)$ in the purely demographic neutral case. Apparently this distribution admits exponential tails that lead, when convoluted with the factor $\exp(-T/\tilde{\tau})$, to moving maximum in the corresponding Laplace integral that yields this value of  $\gamma$.

A similar argument is relevant above the transition zone.  In this case the system jumps between two states, both of them with positive growth rate. The rate of extinction in both states scales like  $\exp(-\alpha N)$, but the coefficient $\alpha$ is  larger in the plus state (linear growth rate $r_0+\sigma$) and smaller in the minus state (linear growth rate $r_0-\sigma$). When extinction happens due to demographic noise, the duration of the extinction event (expected duration of the final decline, \cite{lande2003stochastic}) scales line $\ln N/(r_0 \pm \sigma)$. Accordingly, the dominant route to extinction involves a period $\ln N/(r_0 - \sigma)$ in which the system stays in the minus state. The chance to pick such a period is $N^{-1/\tau(r_0-\sigma)}$, so it contributes only a power-law correction to the $\exp(-\alpha N)$ factor. All in all, in the large $N$ limit the time to extinction in fluctuating environment converges to the time to extinction of the minus state, up to power-law corrections.

\section{Probability distribution function for the time to extinction} \label{pdf}

 Until now we have calculated the mean time to extinction, $T$, in the various phases of the logistic system. Here we would like present a few considerations regarding the full probability distribution function for extinction at $t$, $f(t) dt$, or the survival probability $Q(t) dt$. Of course $$f(t) = -dQ(t)/dt.$$

The state of our system is fully characterized by $P_{e,n}(t)$, the chance that the system admits $n$ A particles  at $t$, when the environmental state is $e$ (for dichotomous noise $e$ takes two values that correspond to $\pm \sigma$). After a single birth-death event (time incremented from $t$ to $t+1/N$), the new state is given by
\begin{equation}
P_{e,n}^{t+1/N} = {\cal M } P_{e',m}^t,
\end{equation}
where ${\cal M }$ is the corresponding Markov matrix. Its matrix elements ${\cal M }_{e,n;e',m}$ are the chance to jump from $m$ particles in environment $e'$ to $n$ particles in environment $e$. These elements  were given in Eqs. (\ref{eqb2}) above.

The highest eigenvalue of the Markov matrix, $\Gamma_0=1$, corresponds to the extinction state, i.e, to the right eigenvector $P_{e,n} = \delta_{n,0}$ (at extinction  the state of the  environment is not significant) or the left eigenvector $(1,1,1,...)$. Using a complete set of left and right eigenvectors of this kind one may write $P_{e,n}(t)$ as,
\begin{equation} \label{sum1}
P_{e,n}(t) = \sum_k a_k v_k (\Gamma_k)^{Nt}.
\end{equation}
Here the index $k$ runs over all eigenstates of the Markov matrix, $v_k$ is the $k$-s right eigenvector, $a_k$ is the projection of $P_{e,n}(t=0)$ on the $k$-s left eigenvector and $Nt$ is the number of elementary birth-death events at time $t$ (for $t=1$, i.e., a generation, $Nt=N$).

Writing $\Gamma_k = |\Gamma_k| \exp(i\phi_k)$, one realizes that each $k$ mode decays like $\exp(-Nt \epsilon_k)$, when $\epsilon_k \equiv -\ln |\Gamma_k|$. Since the Markov matrix is real, eigenvalues are coming in complex conjugate pairs so $P_n$ is kept real and non negative at any time. For the extinction mode $\epsilon_0 =0$, all other modes have  $ \epsilon_k > 0$

Clearly, for any finite system the subdominant mode $\epsilon_1$ determines the maximal persistence time of the system, so at timescales above $t= 1/N \epsilon_1$ the chance of the system to survive, $Q(t)$, decays exponentially with $t$.

\begin{figure}
\includegraphics[width=5cm]{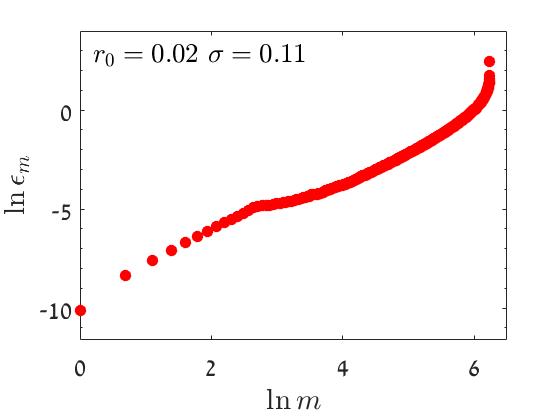}
\includegraphics[width=5cm]{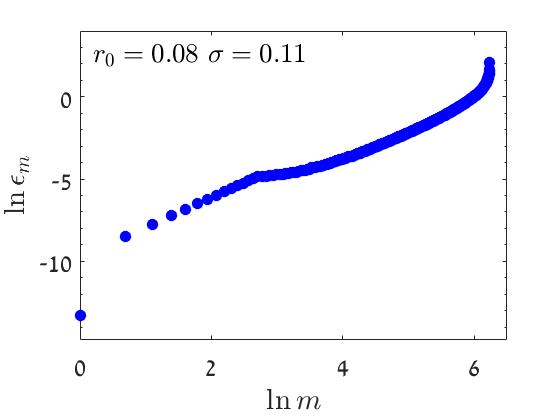}
\includegraphics[width=5cm]{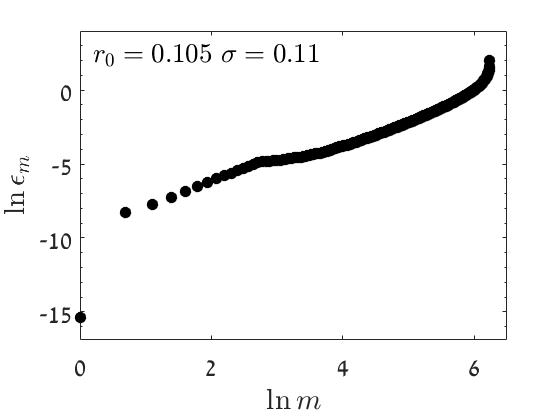}
\caption{The logarithm of the absolute value of the eigenvalues of the Markov matrix, $\epsilon_m$, is plotted against $\ln m$ for small $r_0$ (left panel), intermediate $r_0$ (middle panel) and large $r_0$ (right panel). The state with $m=1$ ($\ln m =0$) is the most persistent  non-extinction state. Clearly, as $r_0$ increases, a gap is opened between $\epsilon_1$ and $\epsilon_2$ (see figure \ref{fig3}). For $m>1$, the low-lying states satisfy $\epsilon_m \sim m^\rho$, where $\rho \approx 1.7$. Parameters are $\tau=1$, $\nu = 0.1$ and $N = 2^8$.  }\label{fig1sup}
\end{figure}

\begin{figure}
\includegraphics[width=5cm]{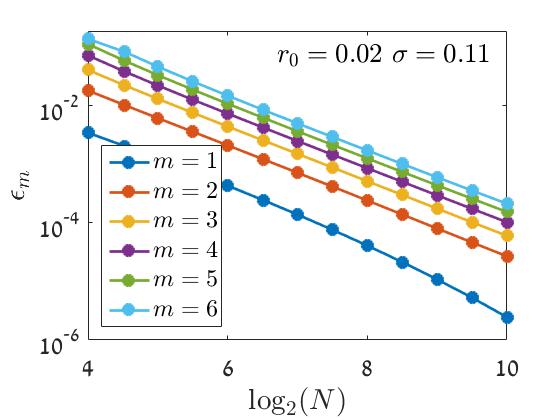}
\includegraphics[width=5cm]{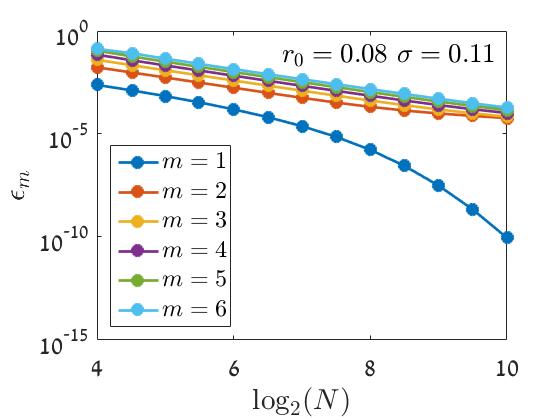}
\includegraphics[width=5cm]{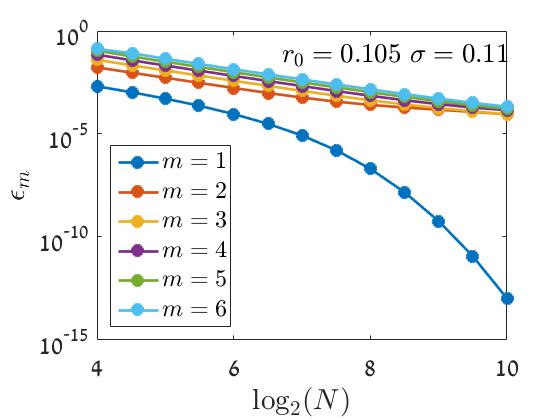}
\caption{ $\epsilon_m$ (from $m=1$ to $m=6$, see legends) is plotted against $\log_2 N$ for different values of $r_0$.  Parameters are $\tau=1$ and $\nu = 0.1$.  }\label{fig2sup}
\end{figure}

\begin{figure}
\includegraphics[width=8cm]{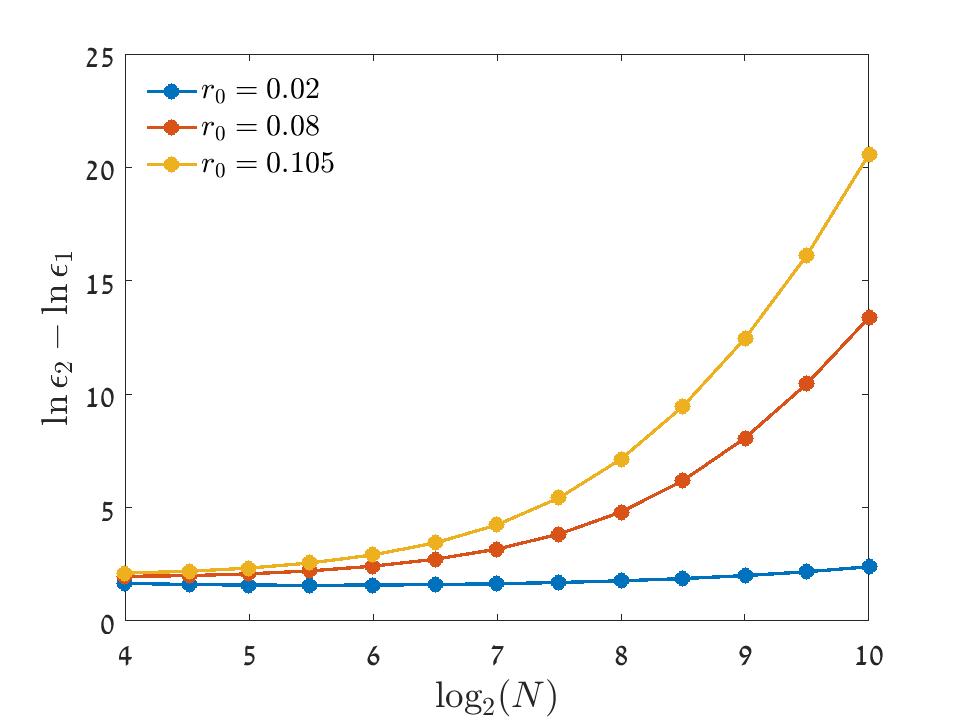}
\caption{ The gap, $\ln \epsilon_1- \ln \epsilon_2$, as a function of $ \log_2 N$. As $N$ increases the gap grows when $r_0$ is large  or intermediate but remains more or less fixed when $r_0$ is small. Parameters are $\tau=1$ and $\nu = 0.1$.   }\label{fig3sup}
\end{figure}

Now one would like to make a distinction between two different situations. In the first, there is a \emph{gap} between $\epsilon_1$ and $\epsilon_2$, so when $N \to \infty$  $\epsilon_1  \ll \epsilon_2 $. This behavior is demonstrated in the right panels of Figures \ref{fig1sup} and \ref{fig2sup} and in Figure \ref{fig3sup}.  In such a case the large $t$ behavior of the system is simply \begin{equation} \label{expo}
Q(t) dt = exp(-t/t_0),
 \end{equation}
where
\begin{equation}
t_0 \equiv 1/N \epsilon_1.
 \end{equation}
 Accordingly, $f(t) = -\dot{Q} =  exp(-t/t_0)/t_0$ and the mean time to extinction $T = t_0 = 1/N \epsilon_1$.

\begin{figure}
\includegraphics[width=5cm]{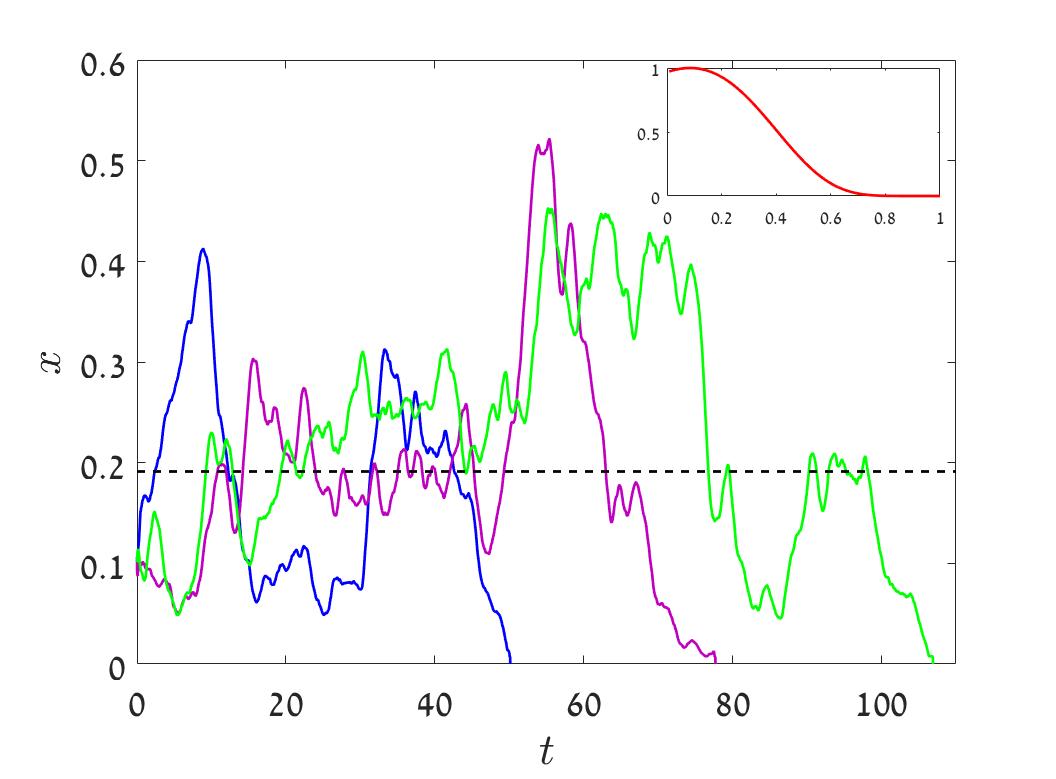}
\includegraphics[width=5cm]{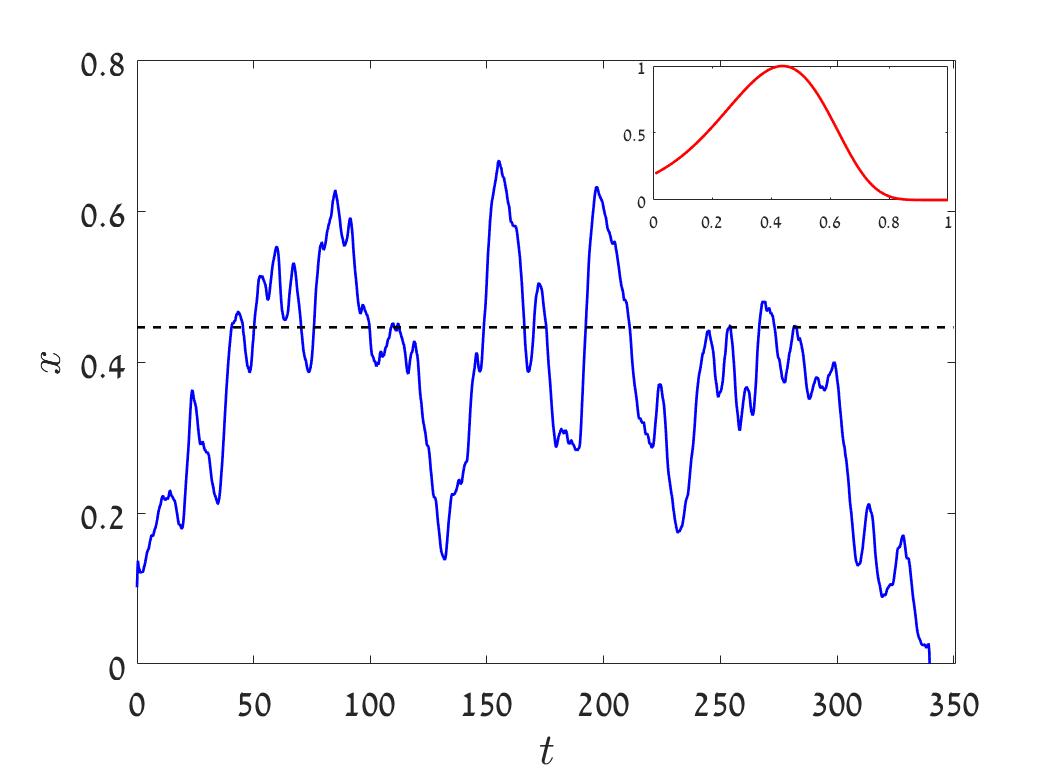}
\includegraphics[width=5cm]{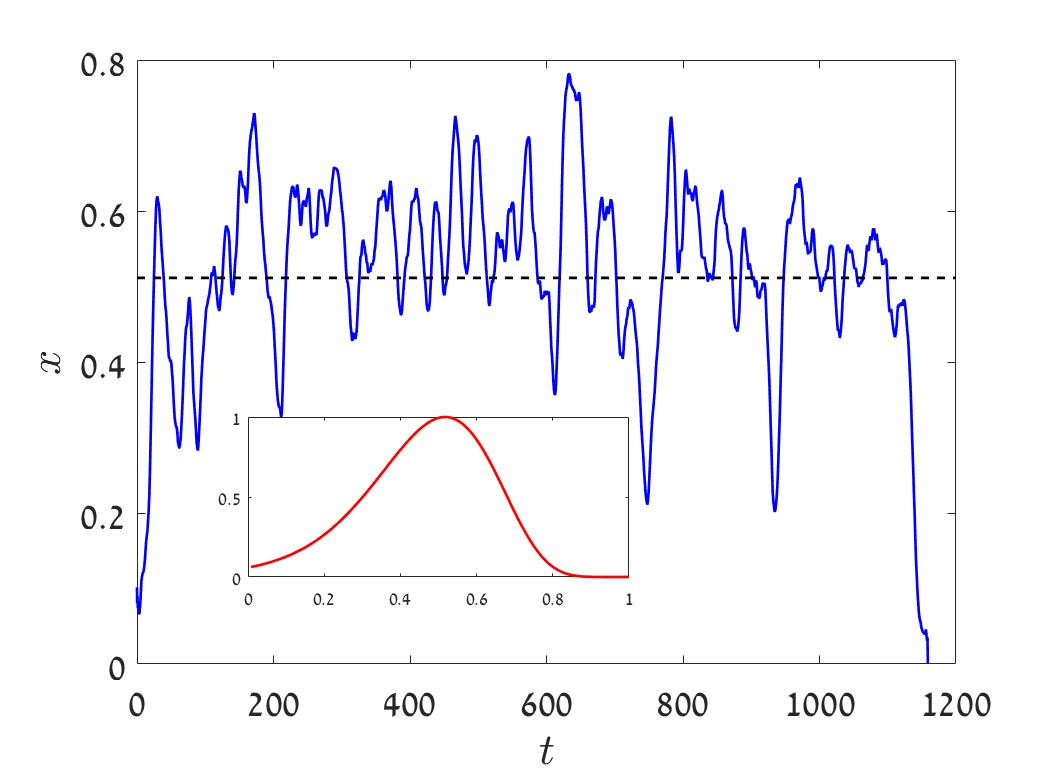}
\caption{ In the main panels, typical trajectories are shown for a system with $\tau=1$, $\sigma =0.11$ and $\nu = 0.1$, for $r=0.02$ (left), $0.08$ (middle) and $0.105$ (right).  The dashed line corresponds to $x^*$, the point where the mean (over environmental conditions) growth rate is zero. In the insets the most stable right eigenstate is plotted for each case. For large $r_0$ the quasistationary state is peaked far from the extinction point, so the trajectory fluctuates in a relatively narrow band around $x^*$. The final decline is sharp: extinction happens due to the accumulation of rare sequences of bad years, and the timescale associated with the  decline time is microscopic (in fact, this timescale is logarithmic in $N$ while the lifetime grows like a power law in $N$). As $r_0$ becomes smaller (the gapless scenario) the fluctuations are comparable with $x^*$, hence the decline time becomes a finite fraction of the lifetime (soft decline).          }\label{figtraj}
\end{figure}

In the exponential phase the situation corresponds to this gap scenario, as discussed in~\cite{kessler2007extinction}. The purely exponential distribution (\ref{expo})  reflects an absence of memory: the system sticks for long times to its quasi stationary state $v_1$, and decay to zero on much shorter timescale due to rare events. The decline to extinction may be a result of a  rare demographic event, like an improbable series of individual death, or  the result of an environmental rare event - an improbable series of bad years. In both cases, the \emph{decline time} (as defined in \cite{lande2003stochastic}) is short (logarithmic in $N$), so the exponential distribution reflects the accumulated chance of rare, short, and independent catastrophes. This \emph{sharp decline} behavior is demonstrated in the right panel of Fig. \ref{figtraj}.

The second scenario (demonstrated in the left panel of Figs \ref{fig1sup} and \ref{fig2sup} and in the blue line of Fig \ref{fig3}) correspond to a \emph{gapless} system. Here the eigenvalues of ${\cal M}$ satisfy $\epsilon_m \sim  \epsilon_1 +c_1 (m-1)^\rho$, where $c_1$ is some tiny constant. In such a case the $\exp(- t N \epsilon_1 )$ factors out of the sum (\ref{sum1}), and the rest of the sum may be approximated by $\int exp(-c_1 t N m^{\rho}) dm$, yielding a power-law decay so,
\begin{equation}
Q(t) dt  \sim \frac{e^{-t/t_0}}{(Nt)^{1/\rho}} dt.
\end{equation} \label{lifetime}
In that case the mean time to extinction is not exactly  $t_0$ but the difference is only a numerical factor. If $\rho>1$ then,
\begin{equation}
T = t_0 (1-1/\rho),
\end{equation}
while if $\rho<1$ the ratio between $T$ and $t_0$ depends on the short time cutoff that must be imposed on the distribution (\ref{lifetime}) to avoid divergence at zero.

This \emph{soft decline} is not purely exponential, since the system has long-term memory. Rare catastrophic events put an upper bound on the lifetime of the population, but extinction may occur, with relatively high probability, due to the random walk of the population size along the log-abundance axis.

In Figure  \ref{figtraj} we show a typical trajectory in each regime, together with a sketch of the right eigenvector (that corresponds to $\epsilon_1$) for the given $r_0$ values. In the gap regime the overlap of the quasi-stationary state with the extinction point is small and a typical trajectory fluctuates around $x^*$ (the point where the mean of $\dot x$ vanishes, where the average is taken over the two signs of $\sigma$), where the amplitude of fluctuations is much smaller than $x^*$. Accordingly, the decline time is relatively short. On the other hand in the gapless case the fluctuation amplitude is larger than $x^*$ and the decline time is comparable with the lifetime.

\section{Universality} \label{universality}

  Through this article we considered one specific microscopic model, namely two species competition with one sided mutation, a classical population-genetics problem taken from \cite{karlin1981second}. Beside its concrete importance, this system is technically more tractable since it corresponds to a zero sum game so the total community size $N$ is strictly fixed and still the system shows negative density dependence.

What about other microscopic processes that yield, in their deterministic limit, a transcritical bifurcation? We would like to suggest that the main features of the transitions considered in this paper are independent of the microscopic details, and are genuine characteristics of transcritical bifurcations under demographic and environmental stocasticity.

In the literature one may find other models that belong to the equivalence class of the logistic growth with environmental stochasticity. These include a model with ceiling (i.e., for which the growth rate is density-independent until it reaches a prescribed value $n^*$, where reflecting boundary conditions are imposed~\cite{lande1993risks,lande2003stochastic}) or simple logistic equation~~\cite{kamenev2008colored,vazquez2011temporal}. Indeed for these models the authors  obtained the same $N$ dependence that we obtained here when the diffusion approximation holds (to the left of the dashed line in Fig. \ref{fig1}). Similarly, the generic Moran process considered in Appendix \ref{deriv} yields Eq. (\ref{eq9}), on which our analysis in the diffusive regime is based.

Moreover, the WKB analysis presented in Section \ref{WKBsec} allows us to suggest a much stronger statement.

As explained, the chance of extinction, and the associated timescale, are given by the behavior of $P(x)$ at $x < 1/N \ll 1$ ($ {\rm Rate} \sim \int_0^{1/N} P(x) \ dx $).
This behavior is determined, in turn, by the small $x$ dependency of $\ell_{+}$ and $\ell_-$. To satisfy Eq. (\ref{trans}) above, $r_0$ must be positive, and this implies that $|\ell_{+}-x|>|\ell_{-}-x|$, i.e., that the mean growth rate of the species, when rare, is positive. This is a sufficient condition for the validity of the WKB analysis presented above, independent of the details of the model at larger $x$-s.

Accordingly, our WKB analysis shows that for \emph{any}  biological species, the time to extinction is a power-law in $N$  when the following conditions are met:
 \begin{itemize}
 	\item The probability distribution function $P(x)$ is normalizable [this condition excludes the logarithmic phase, and ensures that the $N$-dependency comes only from the behavior of $P(x\ll1)$].
 	\item The dynamics allows for periods of growth and periods of decline (this condition excludes the exponential phase).
 	\item When the focal species is rare ($x \ll 1$), its time-averaged growth rate is positive ($|\ell_{+}-x|>|\ell_{-}-x|$).
 	\item When the focal species is rare it grows exponentially  during periods of positive growth rate.
 \end{itemize}

 The last condition excludes cases where the growth or decay when rare are not exponential, e.g., the dynamics of a recessive allele that satisfies $\dot{x} = r(t)x^2$ when $x \ll 1$. In that case the bifurcation is not transcritical.

\section{Practical implications} \label{practical}

The dynamics of a single population is the basic building block of many theories in population genetics, ecology and evolution. As in almost any realistic situation random environmental fluctuations play a major role in system's dynamics, one expects that the phase diagram presented here may shed a new light on the analysis of empirical data in these fields. In this section we would like to sketch of a few emerging insights.

\subsection{Population viability analysis}

Population viability analysis~\cite{lande2003stochastic,lande1993risks,akccakaya2000population,sabo2004efficacy}   is a  method of risk assessment frequently used in conservation biology. Its main aim is to  determine the probability that a population will go extinct within a given number of years. In the typical case the empirical data is abundance timeseries (the number of birds or nests observed at a certain place, the number or the biomass of conspecific trees or shrubs in a region). In some cases these timeseries are collected over a few decade (for example, the North American Breeding Birds survey (NABBS)  has now almost  50 years of large scale censuses).

To extract information, and to suggest predictions based on these timeseries, one would like to analyze them using a decent model. In general a given timeseries is used to infer model parameters and to estimate the strength of stochasticity, then one uses the calibrated model to predict long-term dynamics.

The models used for PVA are almost always logistic or logistic like, so the maximum abundance of a viable population is limited by a certain density-dependent mechanism. The general form of these models is $\dot{x} = R(x)x$, where $R(x)$, the per-capita growth rate,  is a monotonously  decreasing function of $x$ that reaches zero at $x^*$, the stable fixed point. In the logistic model $R(x)$ decays linearly with the population size, in a ceiling model the growth rate is fixed until the population hits the carrying capacity ($R(x)=r_0$ for $x<x^*$ with reflecting boundary conditions at $x^*$), and in Ricker model the birth rate decays exponentially with the population size until it becomes equal to the death rate. The models may differ in some details and may admit discrete time maps ($x_{t+1} = R(x_t)x_t$) but  they all belong to the transcritical bifurcation class. Table 1 of \cite{sabo2004efficacy}, for example, provides references to 27 works in which these models have been used.

However, the predictions of population viability analysis of this kind may be problematic. It is very difficult to estimate the model parameters from empirical timeseries, and on the other hand the results are in many cases very sensitive to these parameters~\cite{ellner2002precision}. The analysis suggested here emphasizes  \emph{universal}, model independent aspects of the dynamics and we would like to suggest our phase diagram as an alternative classification scheme for viability analysis. Instead of trying  to predict the chance of extinction, one would like to adopt a more qualitative approach and to classify populations by their stability properties (the three regimes in Fig. \ref{fig1}) and their decline modes (left/right to the dotted-dashed transition line in that figure).

\begin{figure}[h]
\includegraphics[width=8cm]{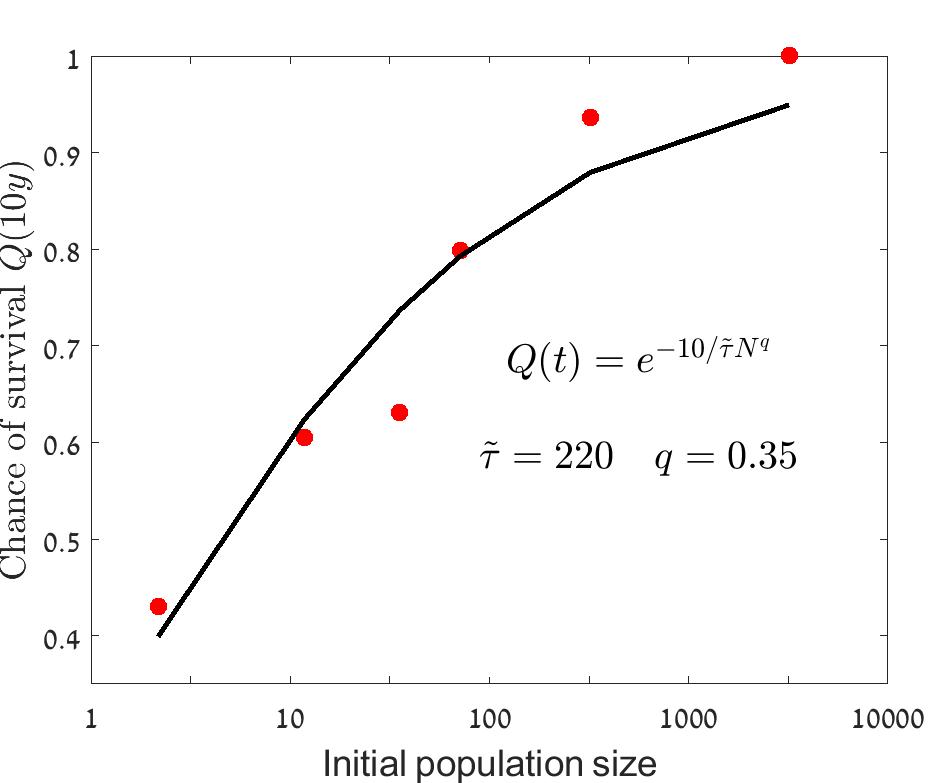}
\includegraphics[width=8.5cm]{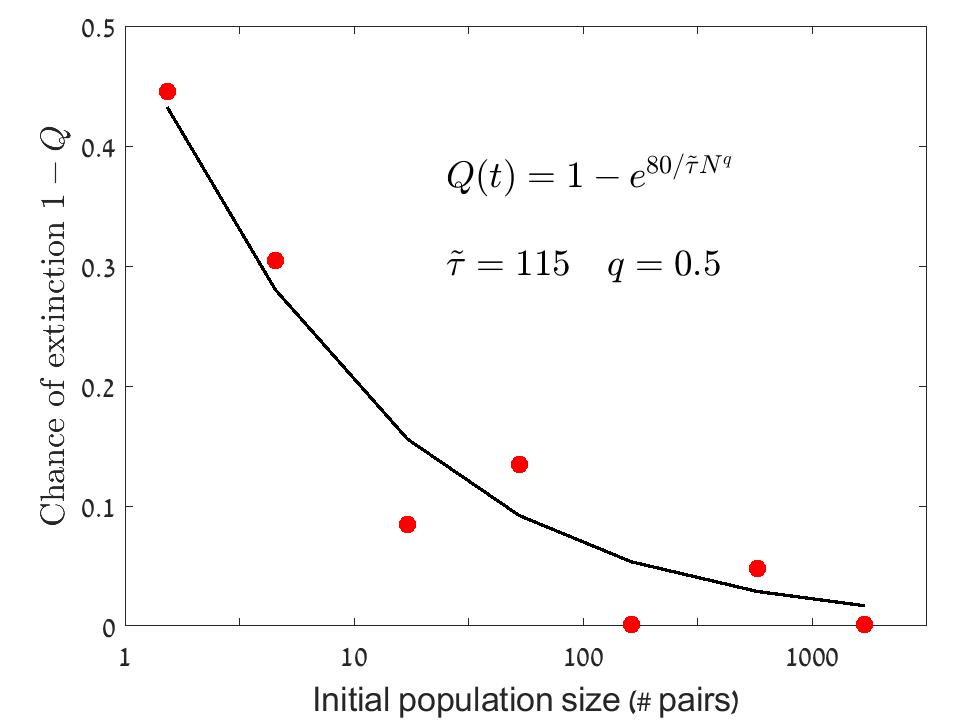}
\caption{The left panel (Figure 1 of \cite{matthies2004population}) shows the relationships between the size of plant populations in 1986 and their chance to survive 10 years later (red circles). The black line is the best fit to $Q(t)$, assuming that the mean time to extinction $T$ growth like  $N^q$. In the right panel we retrieved Figure 5 of \cite{jones1976short}, and the red circles correspond to  the chance of extinction of birds populations vs. the initial number of pairs (the last point in the original figure, that was too close to zero to be digitised, was omitted). The black line is the best fit to $1-Q(t)$, assuming that the mean time to extinction $T$ growth like  $N^q$.  }\label{birds}
\end{figure}

As an example, let us consider the analyses   of~\cite{matthies2004population,jones1976short}. In Figure \ref{birds} we reproduce the relevant datasets from these two papers. In both datasets the chance of survival $Q$, or the chance of extinction $1-Q$, after a fixed time interval (10 or 80 years) are plotted against the initial population size.

As one can see, both datasets (which are, of course, quite noisy because of the small number of samples in each bin, especially for the high abundance bins) allow for reasonable fits if the chance of survival, $Q(t)$, satisfies
\begin{equation}
Q(t) = \exp(-t/\tilde{\tau}N^q),
\end{equation}
 which is the expression one expects if the system is in the power-law phase. Note that the distinction between soft and sharp decline is irrelevant here, since the time window is fixed and we are interested only in the $N$ dependence.

When we tried to fit the data with  $Q = \exp(t/\tilde{\tau}\exp(\alpha N))$, as expected in the exponential phase, we ran into difficulties. In such a case one expects a much steeper dependence of $Q$ on $N$: if $T \sim \exp(\alpha N)$ than when $N$  varies from $0.1/\alpha$ to $10/\alpha$, say,  $Q$ varies from vanishing values to one, so the survival probability  is a sharp sigmoid unless $\alpha$ takes very small values. As a result, we have failed to fit the datasets with exponential dependency unless $\alpha$ was taken to be extremely small ($\alpha \sim 0.01$).  Such a value for $\alpha$ suggests that the mean time to extinction for 100 individuals is just $2.71$ times the mean time to extinction for a single individual, we can not rule out this possibility, but it seems less likely to us.

Moreover, both studies did not report a significant abundance decline in the surviving populations - in most of them abundance either grew up or kept fixed, see Figure 4 of~\cite{jones1976short} and Figure 4 of \cite{matthies2004population}. This implies that both systems are not in the logarithmic phase, where one should expect a general decrease in abundance for all populations.

Accordingly, it seems that a consistent interpretation of the observed data suggests that the surveyed bird and plant populations are in the power-law phase, where the lifetime of a population scales $N^q$, with $q$ values between $0.3$ and $0.5$.

Similar conclusions are suggested by the work of Ferraz et al~\cite{ferraz2003rates}, who measured the rate of bird species loss in amazonian forest fragments as a function of fragment area. If we assume that the initial population size if proportional to the area of the fragment, and that within each fragment the population is more or less well mixed (perhaps a reasonable assumption for birds), the scaling of extinction times with fragment area is the same as the scaling of extinction time with population size. The authors of~\cite{ferraz2003rates} reported that ``A 10-fold decrease in the rate of species loss requires a 1,000-fold
increase in area", suggesting $q \approx 1/3$. However, large fragments may contain more low-density species to begin with, so the actual value of $q$ is perhaps higher than $1/3$.

\subsection{Conservation, management, and decline modes}

The ultimate purpose of conservation efforts is to avoid extinction of local populations. To that aim, we suggest that the distinction between soft decline and sharp decline (the two sides of the dotted-dashed line in Fig. \ref{fig1}) is very important.

As demonstrated in Fig. \ref{figtraj}, in the soft-decline regime (including the logarithmic phase) the system faces a severe risk of extinction \emph{and} its dynamics is strongly related to the  factors that actually lead to extinction, like grazing or fishing or climate change and so on. In this regime one would like to identify these factors and, if possible, to avoid them.

The situation is completely different in the sharp decline regime, where the system fluctuates for a long time around an equilibrium value and extinction is caused by rare events (a long series of bad years, for example). Predicting rare events from a given, relatively short, timeseries is usually a lost cause, so one would like to direct conservation efforts in other directions.

Decline modes may be inferred from the statistics of extinction events in timeseries, provided that false extinction events (due to  sampling errors) are filtered out somehow.  Large scale empirical studies of $Q(t)$ (like those presented in \cite{keitt1998dynamics,bertuzzo2011spatial}, based on the NABBS data) suggest an exponentially truncated power law. If one likes to interpret these results as reflecting purely local logistic-like dynamics under environmental stochasticity (this was not the interpretation given in~\cite{bertuzzo2011spatial} - they considered a neutral model with immigration), it implies that the decline mode in these systems is indeed soft.

\subsection{Spatial effects}

One of the main obstacles to the understanding of ecological dynamics is the need to bridge over length and time scales. Understanding of stability developed at small scales cannot be easily extended to larger scales, since the type and effect of ecological  processes  vary  with  scale.  For example, the work of~\cite{ferraz2003rates} suggests, as mentioned above, that  a ten fold decrease in  $t_{1/2}$ (time to lose $50 \%$ of the species) requires a 1000 fold increase in fragment size, and that $t_{1/2}$   for a fragment of 100 ${\rm km}^2$ is 100 years. This observation is based on a study of fragments between 0.01 and 100 ${\rm km}^2$. Extrapolating this scaling law to the whole tropical south America subregion ($1.4 \cdot 10^7$ ${\rm km}^2$) yields a mean lifetime of about 10,000 years for a tropical bird species, much shorter that the standard estimates from fossil data (about a million years per species).

Our discussion so far points towards the importance of the correlation length, $\xi_E$ associated with environmental fluctuations. When the linear size of a system is much larger than this correlation length local extinctions are uncorrelated and may be compensated by recolonization of empty patches by neighboring populations (a metapopulation dynamics~\cite{hanski1998metapopulation}). In that case the extinction-recolonization dynamics is effectively a contact process, and the  extinction transition belongs to the directed percolation equivalence class~\cite{hinrichsen2000non}. Technically speaking, for  the directed percolation transition spatio-temporal noise is an irrelevant operator that does not affect the transition exponents.

On the other hand, when $\xi_E$ much larger than the linear size of the system temporal stochasticity is a relevant operator and the transition belongs to a different equivalence class~\cite{barghathi2017extinction}.  In that case a series of bad years may kill the whole system and the chance of recolonization is small since the state of  all neighboring sites is correlated. Accordingly, one expects  that an increase in $\xi_E$ makes the system less stable~\cite{liebhold2004spatial}. Upscaling of local observation is perhaps limited to length scales below $\xi_E$.

\subsection{Existence and coexistence}

The modern coexistence theory  (MCT) have gained a lot of attention in recent years  \cite{ellner2018expanded,barabas2018chesson}. The aim of this theory is to understand the conditions for coexistence of many species.  Here we considered a system with only two species (and effectively one, since the game is zero-sum), still our results are relevant to one aspect (out of many) of the coexistence theory.

In MCT  one analyzes the dynamics of a single species in an ``effective field" that reflects both the environment and the interaction with  all other species~\cite{schreiber2012persistence}. Demographic noise is not taken into account and the ``existence" of a single species is related to the chance of its abundance to visit the extinction zone, which is the region between $x=0$  and $x= \delta$. Clearly a reasonable value for $\delta$ is $1/N$ (as in Section \ref{WKBsec} above), but the theory has no demographic stochasticity in it so $\delta$ is left arbitrary. A species coexists if, for every $\epsilon>0$ there exists a $\delta$ such that the system spends less than $\epsilon$ of its lifetime in the extinction zone~\cite{schreiber2012persistence}.

Our work shows that this is a very weak statement. For our system, the MCT persistence criteria is translated to $r_0>0$ (alternatively, to the divergence of $T$ when $N \to \infty$), but in this regime the time to extinction scales like $N^{r_0/g}$ and the exponent may take arbitrary small values ($N^{0.01}$, say), so in any realistic system  the focal species will go extinct in a very short time. We believe that the distinction between sublinear and superlinear dependencies of $T$ on $N$, or the decline mode, are much more important than the condition for ``coexistence".

\section{Discussion}

Through this paper we considered mainly the large-$N$ asymptotics, under the assumption that all other parameters are kept fixed and $N$ is taken to infinity. In that case, the system allows for an exponential phase only when the noise is \emph{bounded}. If the noise is unbounded (e.g., Gaussian noise) these is always a finite chance to pick a long series of bad years (a long period of time in which the linear growth rate is negative). This chance may be extremely tiny, but it is  $N$-independent. As a result, it dominates the large $N$ limit so in that limit one finds only two phases, the logarithmic phase and the power-law phase.

Practically, the distinction between a power-law phase with diverging $q$ and an exponential phase is, in almost any imaginable case, unimportant. We believe that in most cases the transitions, or crossovers, associated with the dotted-dashed line of Fig. \ref{fig1} are much more relevant to the analysis of empirical dynamics.

Three  transitions/crossovers may be identified within the power-law phase. The first has to do with the value of $r_0/g$. When this value is smaller than one the time to extinction is sublinear in $N$ while above this point it is superlinear \cite{spanio2017impact}. The  corresponding $P(x)$ (that may be obtained from $P(y)$ that was calculated in section \ref{WKBsec} using the appropriate Jacobian $1/x$), behaves, close to zero, like $P(x \ll 1) \sim x^{-1+r_0/g}$, so $P(x)$ changes its shape from convex to concave.

The second transition is the breakdown of the diffusion (continuum) approximation. This has to do with the dependency of $P_n$ on $n$ in the discrete master equation (\ref{eqb1}). When this dependency is too steep, such that gradients decay only slowly with $N$, the continuum approximation fails and one has to implement WKB.

A third phenomenon, discussed in Section \ref{pdf}, is the opening of a gap in the spectrum of the Markov operator as $r_0$ increases. As we have seen, in a system with a gap the decline is sharp and the dynamics has no memory: during every short period of time (of order $\ln N$) the process either dies or stay alive, and persistence times have an exponential distribution. When the gap closes down the probability distribution function follows a  truncated power law and extinction occurs because of the random motion along the abundance axis (soft decline).

One may wonder about the relationships between these three transitions. When $P(x)$ changes its shape from convex to concave, the fraction of time that a typical trajectory spends in the extinction zone shrinks more rapidly with $N$, so extinction becomes more and more associated with rare events, and this suggests that the spectrum admits a gap. Similarly, when extinction happens due to rare fluctuations associated with directed flow towards zero, it is quite plausible that the diffusion approximation breaks down, since the convergence of a binomial distribution to a Gaussian is known to fail at the far tails of the distribution. Our analysis so far do not allow us to declare that these three phenomena are all manifestations of the same transition, more work is still needed.

We acknowledge many helpful discussions with David Kessler and Matthieu Barbier.  This research  was supported by the ISF-NRF Singapore joint research program (grant number 2669/17).

\bibliography{refs}

\clearpage

\appendix

\section{Derivation of Eq. \ref{eq9}} \label{deriv}

In this appendix, we present two derivations of Eq. (\ref{eq9}) above. The first derivation is based on the specific local competition model we implemented here, as described in Section \ref{model}. The second derivation is based on a generic, continuous time Moran process. As one shall see, the first derivation more complicated and less generic, however it is needed since our numerical analyses are based on Eq. (\ref{eqb1}).

\subsection{Derivation using the local competition model}

In the process used through this paper, in each elementary step two individuals are chosen at random for a duel. In the main text we defined the chance of intraspecific and interspecific competition for the focal species, $F_n$ and $Q_n$, and the chance of the focal species to win an interspecific competition event in the plus and the minus state, $P_A^{\pm}$.  Given these specifications of the model,  the transition probabilities are:
\begin{eqnarray} \label{eqb2}
W^{++}_{n \to n-1} = \left(1-\frac{1}{\tau N}\right)  \left[ F_{n} (1-P_A^+) + \nu Q_n \right] &\qquad& W^{++}_{n \to n+1} = \left(1-\frac{1}{\tau N}\right)  \left[ (1-\nu) F_{n} P_A^+  \right]  \\
W^{--}_{n \to n-1} = \left(1-\frac{1}{\tau N}\right)  \left[ F_{n} (1-P_A^-) + \nu Q_n \right] &\qquad& W^{--}_{n \to n+1} = \left(1-\frac{1}{\tau N}\right)  \left[ (1-\nu) F_{n} P_A^-  \right] \nonumber \\
W^{-+}_{n \to n-1} = \frac{1}{\tau N}  \left[ F_{n} (1-P_A^+) + \nu Q_n \right] &\qquad& W^{-+}_{n \to n+1} = \frac{1}{\tau N} \left[ (1-\nu) F_{n} P_A^+  \right] \nonumber \\
W^{+-}_{n \to n-1} = \frac{1}{\tau N}  \left[ F_{n} (1-P_A^-) + \nu Q_n \right] &\qquad& W^{+-}_{n \to n+1} = \frac{1}{\tau N} \left[ (1-\nu) F_{n} P_A^-  \right]. \nonumber
\end{eqnarray}
The superscripts of the $W$-s refer to the environmental state, where switches take place before the competition step, so for example $W^{+-}_{n \to n-1}$ is the probability that the environment switched from the plus to the minus state and then the focal species lost a single individual.

The mean time to extinction for species A, when it is represented by $n$ individuals in the plus state, $T^+_n$, and the corresponding quantity in the minus state, $T^-_n$, satisfy the discrete backward Kolmogorov equation (BKE),
\begin{eqnarray} \label{eqb1}
T_{n}^+ &=&   W^{++}_{n \to n+1} T_{n+1}^+  + W^{++}_{n \to n-1} T_{n-1}^+ +(1-W^{++}_{n \to n+1}-W^{++}_{n \to n-1}) T_{n}^+    \nonumber \\ &+& W^{+-}_{n \to n+1} T_{n+1}^-  + W^{+-}_{n \to n-1} T_{n-1}^- +(1-W^{+-}_{n \to n+1}-W^{+-}_{n \to n-1}) T_{n}^-   +\frac{1}{N} \\
T_{n}^- &=&   W^{--}_{n \to n+1} T_{n+1}^-  + W^{--}_{n \to n-1} T_{n-1}^- +(1 -W^{--}_{n \to n+1}-W^{--}_{n \to n-1}) T_{n}^-   \nonumber \\ &+& W^{-+}_{n \to n+1} T_{n+1}^+  + W^{-+}_{n \to n-1} T_{n-1}^+ +(1 -W^{-+}_{n \to n+1}-W^{-+}_{n \to n-1}) T_{n}^+  +\frac{1}{N}. \nonumber
\end{eqnarray}
The boundary conditions at the absorbing state are $T_0^+ = T_0^-=0$. In the other end $n=N$ the boundary conditions are determined by the relationships (imposed by the transition probabilities)  between $T^\pm_{N-1}$ and $T^\pm_N$ (see below).

Using a linear transformation, $$T_n \equiv (T^+_n + T^-_n)/2 \qquad \Delta_n \equiv (T^+_n - T^-_n)/2,$$ one may rewrite (\ref{eqb1}) in a more informative manner, since $T_n$ is the mean time to extinction where the average is taken over both histories and initial state of the environment.
In the limit $N \gg 1$  the  continuum approximation is used. $n/N$ is replaced by $x$, $n \pm 1  \equiv x \pm 1/N$,  and all the functions $T$ and $W$ are expanded to second order in $1/N$. The resulting set of equations is,
\begin{eqnarray} \label{eq7}
\left(s_0 (1-\nu/2) - \frac{\nu}{1-x} \right)T'(x) + \left(1+\frac{\nu}{2} \left[\frac{ 2x-1}{1-x} -\frac{s_0}{2} \right] \right) \frac{T''(x)}{N} +\tilde{\sigma}(1-\nu/2)\Delta'(x) - \frac{\tilde{\sigma} \nu \Delta''(x)}{4N} &=& -\frac{1}{x(1-x)} \\
\left(s_0 (1-\nu/2) - \frac{\nu}{1-x} \right)\Delta'(x) + \left(1+\frac{\nu}{2} \left[\frac{ 2x-1}{1-x} -\frac{s_0}{2} \right] \right) \frac{\Delta''(x)}{N} +\tilde{\sigma}(1-\nu/2)T'(x) - \frac{\tilde{\sigma} \nu T''(x)}{4N} &=& \frac{2N}{N\tau-2} \frac{\Delta}{ x(1-x)} \nonumber.
\end{eqnarray}

To proceed, we make the following steps,
\begin{enumerate}
  \item We assume $N\tau \gg1$, so $N \tau -2 \approx N\tau$. This implies that as $N$ is taken to be large $\tau$ is kept finite, i.e., that the persistence time of the environment is independent of the size of the community.
  \item In the second Equation of (\ref{eq7}) only the $\Delta$ and the $T'$ terms are kept. In general, applying a dominant balance argument one finds two dominant terms in the large $N$ limit. One of them must be the $\Delta$ term, otherwise the result is independent of the persistence time of the environment $\tau$, which is physically impossible. This term must be balanced by one of the $T$ terms (if this is not the case $\Delta =0$ is a solution and environmental stochasticity has no effect), and the continuum approximation is valid only if $T'>>T''/N$.  Using that we can solve for $\Delta$ and  plug the solution into the upper equation of (\ref{eq7}).
      \item The $\Delta''$ term in the upper equation is neglected, again this holds when the continuum approximation is applicable.
\end{enumerate}
Under these approximations, Eqs (\ref{eq7}) reduce to a single, second order, inhomogeneous differential equation for $T$,
\begin{eqnarray}\label{eq8}
\left(s_{0} (1-\nu/2) - \frac{\nu}{1-x} + \frac{[\tilde{\sigma}(1-\nu/2)]^2 \tau}{2}(1-2x) \right)T'(x) &+& \\ \nonumber   \left(1+\frac{\nu}{2} \left[\frac{ 2x-1}{1-x} -\frac{s_{0}}{2} \right] + \frac{N[ \tilde{\sigma}(1-\nu/2)]^2 \tau}{2} x (1-x) \right) \frac{T''(x)}{N}  &=& -\frac{1}{x(1-x)}.
\end{eqnarray}

Assuming further $\nu \ll 1$ and $s_0 \ll 1$ (this implies that the deterministic growth/decay during one generation is small with resect to the population size), Eq. (\ref{eq8}), with the definition $g \equiv \sigma^2 \tau/2$, is reduced to Eq. (\ref{eq9}) of the main text.

The boundary conditions for Eq. (\ref{eq9}) are derived from the discrete equations in the appropriate limit. Since $T^+(0)=T^-(0)=0$, in the continuum limit $T(0)=0$. In the reflecting boundary $x=1$ one may use the discrete equations
\begin{eqnarray} \label{eqf}
T^+_{N} = (1-1/N\tau) [(1-\nu)T^+_{N}+\nu T^+_{N-1}] +(1/N\tau) [(1-\nu)T^-_{N}+\nu T^-_{N-1}] +1/N \nonumber \\
T^-_{N} = (1-1/N\tau) [(1-\nu)T^-_{N}+\nu T^-_{N-1}] +(1/N\tau) [(1-\nu)T^+_{N}+\nu T^+_{N-1}] +1/N \end{eqnarray}
so
\begin{equation}
T_N = (T^+_{N}+T^-_{N})/2 = (1-\nu)T_{N}+\nu T_{N-1} +1/N.
\end{equation}
Accordingly,
$$ T'(1) = \frac{1}{\nu}.$$

\subsection{Derivation for a generic Moran process}

Here we would like to derive Eq. (\ref{eq9}) for a simple birth-death-mutation  process which is defined for a single species  by the rates of the demographic and the environmental events (a continuous time  Moran process). In this section we consider environmental transitions between different $s$ states, where $s$ determines the birth and the death rates. For a given $s$, the rates are,
\begin{eqnarray}
W_{1,s} =  W(x,s \to  x+1/N,s) &=& N x (1-x) (1+s) \nonumber \\
W_{2,s} =  W(x,s \to  x-1/N,s) &=& N x (1-x) (1-s) + N \nu x \nonumber \\
W_3 = W(x,s \to  x,s') &=& 1/\tau.
\end{eqnarray}

The $W$s were chosen such that the total rate of birth and death events does not depend on the environment $s$, in agreement with the previous model. If this is not the case the number of birth and death events between two environmental shifts depends on $s$, so the amplitude of fluctuations in population size per unit time depends on the environmental conditions. Although the effect will be small for $s \ll 1$, we would like to avoid the corresponding  redundant terms in the analysis.

To allow a comparison with the derivation in the last subsection we translate rates to probability by dividing the relevant value of $W$ by $W_T = W_1+W_2+W_3$  (Note that $W_T$ is independent of $s$ by construction). The BKE for the two branches then read,

\begin{equation}
T_{n,s} = \frac{W_{1,s}}{W_T}T_{n+1,s}+\frac{W_{2,s}}{W_T}T_{n-1,s}+ \frac{W_3}{W_T} T_{n,s'} + 1/W_T.
\end{equation}

In the continuum limit  ($T_{n \pm 1}$ as $T(x)+T'(x)/N + T''(x)/(2N^2)$) one finds,
\begin{equation}
 \frac{\mu(x,s)}{N} T' +\frac{\Sigma(s,x)}{2N^2} T'' + W_3[T(s',x)-T(s,x)]  =-1
\end{equation}
where $\mu(s,x) \equiv W_{1,s}-W_{2,s}$ and $\Sigma(s,x) \equiv W_{1,s}+W_{2,s}$.

Defining $T = [T(s_1) + T(s_2)]/2$ and $\Delta = [T(s_1) - T(s_2)]/2$ one finds,
\begin{eqnarray} \label{eqz}
(\mu_1+\mu_2) T' + \frac{1}{2N} (\Sigma_1 + \Sigma_2) T'' + (\mu_1-\mu_2) \Delta'  +\frac{1}{2N} (\Sigma_1 - \Sigma_2) \Delta'' = -2N \nonumber \\
(\mu_1+\mu_2) \Delta' + \frac{1}{2N} (\Sigma_1 + \Sigma_2) \Delta'' + (\mu_1-\mu_2) T'  +\frac{1}{2N} (\Sigma_1 - \Sigma_2) T'' = \frac{2N}{\tau}\Delta.
\end{eqnarray}

If $s_1 = s_0 + \sigma$ and $s_2 = s_0 - \sigma$, $\mu_1 = 2Nx(1-x)(s_0+\sigma)-N\nu x$, $\mu_2 = 2Nx(1-x)(s_0-\sigma)-N\nu x$, and $\Sigma_1 = \Sigma_2 = 4Nx(1-x)+2N \nu x$.

In the second equation of  (\ref{eqz}) we keep only the $\Delta$ and the $T'$ terms, and obtain from them an expression for $\Delta''$ (Note that $\Sigma_1 - \Sigma_2 =0$ so the $T''$ term in the second equation and the $\Delta''$ term in the first one vanishes). Plugging $\Delta'$ in the first equation one gets Eq. (\ref{eq9}) of the main text.

\section{Numerical methods} \label{methods}

Through this work we compare results, obtained from numerical solutions of the backward Kolomogorov equations (BKE), to the our analytic approximations.

\subsection{Numerical solution of the backward Kolomogorov equation}

The discrete BKEs considered through this paper, (Eq. \ref{eqb1}) is second order, linear, inhomogeneous difference equations that have the general form,
\begin{equation}
\left[ \begin{array}{c} T^+_1 \\ \vdots \\ T^+_{N-1} \\ T^-_1 \\ \vdots \\ T^-_{N-1}   \end{array} \right] = \begin{bmatrix} W^{++}_{1 \to 1} &  W^{++}_{1 \to 2} & \cdots &  W^{+-}_{1 \to 1} &  W^{+-}_{1 \to 2} & \cdots \\  \vdots & &\ddots & &\vdots \\  \vdots & &\ddots & &\vdots\\  W^{-+}_{1 \to 1} &   W^{-+}_{1 \to 2} & \cdots &  W^{--}_{1 \to 1} &  W^{--}_{1 \to 2} & \cdots \\  \vdots & &\ddots & &\vdots \\  \vdots & &\ddots & &\vdots  \end{bmatrix} \times \left[ \begin{array}{c} T^+_1 \\ \vdots \\ T^+_{N-1} \\ T^-_1 \\ \vdots \\ T^-_{N-1}   \end{array} \right]+\frac{1}{N} \left[ \begin{array}{c} 1\\ \vdots \\ \vdots \\ \vdots \\ \vdots \\ 1 \end{array} \right],
\end{equation}
with the $W$-s that were defined in Eq. (\ref{eqb1}) above.  Accordingly, the values of $T^{\pm}_n$ [and consequently the values of $T_n$ and $\Delta_n$] may be determined by inverting this $(2N-2) \times (2N-2)$ matrix and multiplying the outcome by the constant vector $-1/N$.

However, in the deep power-law pase and in the exponential phase this procedure suffers from numerical errors, as one can see in Fig. (\ref{continue}). To test our WKB analysis we had to overcome this difficulty and to increase the machine precision.  Wolfram's \emph{Mathematica} provides a tempting opportunity, as it allows one to work with infinite precision variables, but is limited by its ability to invert large matrices efficiently.

To solve this problem we have implemented our numerics in \emph{Mathematica} using a transfer matrix approach. Eq.  (\ref{eqb1}) may be written as

\begin{equation}
\left[ \begin{array}{c} T^-_n\\ \\ T^+_n \\ \\ T^-_{n+1} \\ \\ T^+_{n+1} \\ \\ 1/N \end{array} \right] =  M_n \times \left[ \begin{array}{c} T^-_{n-1}  \\ \\ T^-_{n-1} \\ \\ T^-_n \\ \\  T^+_{n} \\  \\ 1/N  \end{array} \right],
\end{equation}
where $M_n$ is  the transfer matrix,
\begin{equation} M_n \equiv \begin{bmatrix} 0 & 0 & 1 & 0 & 0\\  \\ 0 & 0 & 0 & 1 & 0 \\ \\-\frac{W^{--}_{n \to n-1}}{W^{--}_{n \to n+1}} & 0 &1+\frac{W^{--}_{n \to n-1}+(\tau N -1)/\left(\tau N (\tau N -2)\right)}{W^{--}_{n \to n+1}} & \frac{1- \tau N}{\tau N (\tau N -2)W^{--}_{n \to n+1}} & \frac{1-\tau N}{\tau N W^{--}_{n \to n+1}}\\ \\ 0 &-\frac{W^{++}_{n \to n-1}}{W^{++}_{n \to n+1}}  & \frac{1- \tau N}{\tau N(\tau N -2)W^{++}_{n \to n+1}} & 1+\frac{W^{++}_{n \to n-1}+(\tau N -1)/\left(\tau N (\tau N -2)\right)}{W^{++}_{n \to n+1}} & \frac{1-\tau N}{\tau N W^{++}_{n \to n+1}}\\ \\ 0 & 0 & 0 & 0 & 1  \end{bmatrix}.
\end{equation}

$T$ may be incremented by $n+1$ abundance steps by multiplication of such matrices,
\begin{equation}
\left[ \begin{array}{c} T^-_n\\ \\ T^+_n \\ \\ T^-_{n+1} \\ \\ T^+_{n+1} \\ \\ 1/N \end{array} \right] =  M_{n,m} \times \left[ \begin{array}{c} T^-_{m-1}  \\ \\ T^-_{m-1} \\ \\ T^-_m \\ \\  T^+_{m} \\  \\ 1/N  \end{array} \right].
\end{equation}
where $M_{n,m} \equiv M_n \times M_{n-1} \times \cdots \times M_{m+1} \times M_{m}$ (assuming $n>m$).

  The transfer matrix $M_{N-1,1}$ may be used to find $T^\pm_N, T^\pm_{N-1}$ as function of $T^+_1, T^-_1$, starting from the column vector $[0,0,T^-_1,T^+_1,1/N]$.  The boundary condition equation at $N$  (\ref{eqf}) provide us with another pair of equations,
\begin{eqnarray}
\frac{T^+_{N}-T^+_{N-1}}{2} + \frac{T^-_{N}-T^-_{N-1}}{2} = \frac{1}{\nu} \nonumber \\
\frac{T^+_N-T^-_N}{T^+_{N-1}-T^-_{N-1}} = \frac{\nu}{\nu+2/(\tau-2)},
\end{eqnarray}
which allows one to solve for $T^+_1, T^-_1$ and to find $T^\pm_n$ for any $n$ by multiplying the column vector $[0,0,T^-_1,T^+_1,1/N]$ by  $M_{n-1,1}$.

This way of using transfer matrix allows  one to find $T$ for big systems in high accuracy where only the values of the $5 \time 5$ transfer matrix are kept in the memory of the system.

However,  because \emph{Mathematica} adapts to the number of digits in the relevant calculation, one would like to avoid the multiplication of matrices one by one, since the number of digits in each element of the matrix increases and this consumes a lot of computer time. To allow for faster calculations, we have generated first all the transfer matrices $M_1 ,M_2, \cdots M_{N-2}, M_{N-1}$, then multiply all pairs of adjacent matrices and repeat the process ${\cal O} (\ln N)$ times to obtain $M_{n-1,1}$.

 \section{Dichotomous (telegraphic) and other types of noise} \label{telegraphic}

 In this article we  consider a special type of environmental stochasticity, in which the system flips between two states (good and bad years, say). Both white Gaussian noise and white Poisson noise can be recovered from this dichotomous (telegraphic) noise by taking suitable limits~\cite{ridolfi2011noise}, so the results obtained here are quite generic.

 As an example, if the environmental conditions are picked from a Gaussian distribution of a certain width with correlation time $\tau_1$, one may easily imitate these features by taking a dichotomous noise that flips between two values, $\pm \sigma$, with much shorter correlation time $\tau$. With the appropriate choice of $\tau$ and $\sigma$, the binomial distribution of $\sigma_{eff}$,  the average fitness between $0<t<\tau_1$,
 \begin{equation}
 \sigma_{eff} = \frac{\tau_1}{\tau} \sum_i^{\tau_1/\tau} \sigma_i,
 \end{equation}
 will correspond to the bulk properties of any required Gaussian noise, since the Gaussian distribution is the limit of a binomial distribution.

 However, while the Gaussian distribution is unbounded, the distribution of $\sigma_{eff}$ is clearly bounded; the convergence  to a Gaussian takes place in the bulk but  the tails are truncated.

\begin{figure}[h]
\includegraphics[width=8cm]{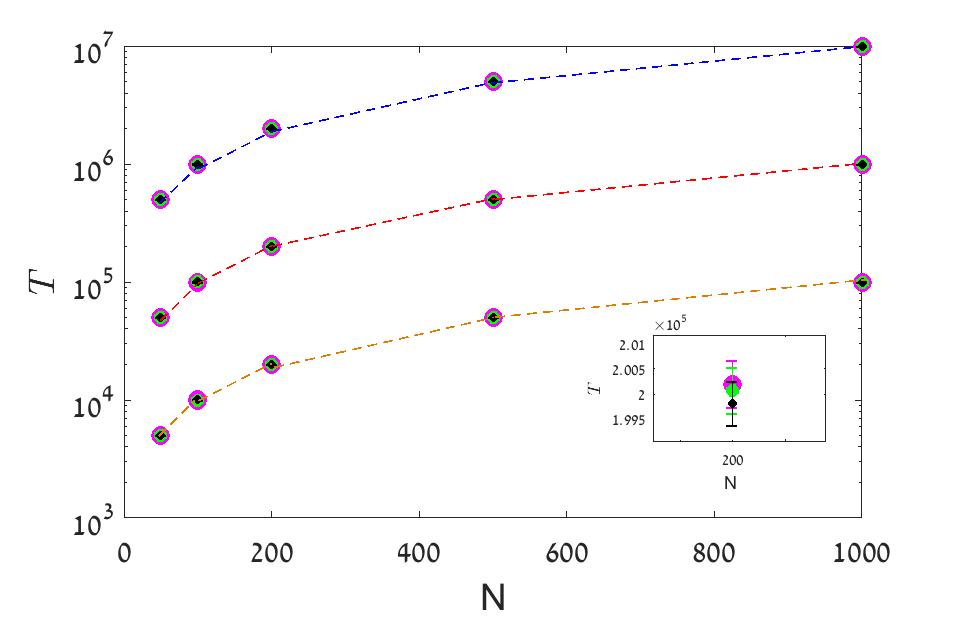}
\includegraphics[width=8cm]{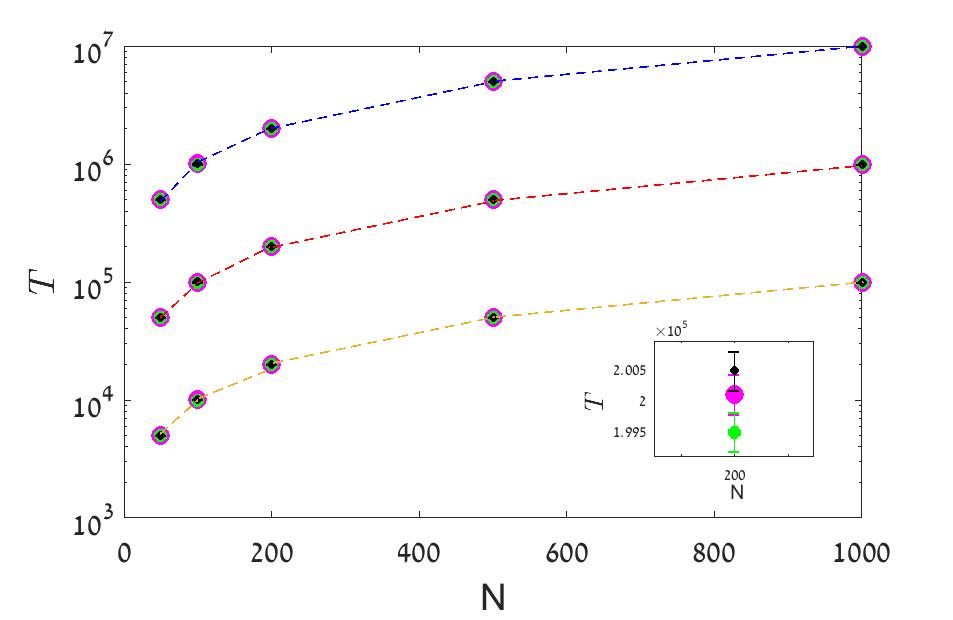}
\caption{Time to extinction $T$ (log scale) vs. $N$ for three different noise distributions. The mean (over 1000-2000 runs) time to extinction was measured as a function of $N=50,100,200,500,1000$, for $n_0=N$. The left panel present results for $\tau= \sigma = 0.1$ while in the right panel $\tau=\sigma=0.3$. For each $N$ and $\nu$ the value of $T$ is given for dichotomous noise (green circles), uniform distribution (magenta) and Beta distribution (black). Markers were chosen with different size to improve the visibility of the results. Dashed line were added manually to guide the eye and they connect results with $\nu = 0.01$ (yellow) $\nu=0.001$ (red) and $\nu = 0.0001$ (blue). In the insets the three points  at $N=200$, $\nu = 0.001$, with one standard deviation error bars, were magnified.  These error bars are too small and cannot be seen in the main panels.   }\label{simu}
\end{figure}

 To demonstrate the ability of a dichotomous noise to emulate the effect of other types of noise, we present in Figure \ref{simu} the outcomes of a few numerical experiments. The figures show the mean time to extinction vs. $N$ for our two-species competition model with one sided mutation, as described in the main text [Eq. (2)].  Three types of noise are compared.
 \begin{enumerate}
   \item $s(t)$ is either $\sigma$ or $-\sigma$ (dichotomous noise).
   \item $s(t)$ is picked from a uniform distribution between $(-\sigma \sqrt{3})$ and $(+\sigma \sqrt{3})$.
   \item $s(t)$  is picked from a beta distribution, $ \rm{Beta}(2,2)\sigma/\sqrt{0.05}$.
 \end{enumerate}
 All three distribution have a compact support, zero mean and variance $\sigma^2$.

\section{The typical path to extinction} \label{path}

Through this paper we have considered for a few times the typical path to extinction when $N$ is large and $r_0 \le 0$. In particular in the power law phase we assumed that when the bias is essentially towards larger population, the most probable path to extinction is a rare series of jumps towards extinction. During a bad year the population decays exponentially, so the minimal length of such a series is ${\cal O} (\ln N)$. We assumed that ${\cal O} (\ln N)$ sequence  is more probable than longer sequences, although the number of possible trajectories growth exponentially with their length.

To substantiate this argument, let us consider the simple case of a random walker on a semi-infinite line, where the walk starts at $n_0$ and in each step the walker jumps to the right with probability $p$ and to the left with probability $q=1-p$, where we assume $p>1/2$. To reach extinction in $S$ steps the trajectory has to include $S/2+n_0/2$ steps to the left and $S/2-n_0/2$ to the right. Accordingly, the probability of  extinction in $S$ steps is,
\begin{equation}
 K = {S \choose S/2+n_0/2 } p^{S/2-n_0/2} q^{S/2+n_0/2}.
\end{equation}
Using Stirling's approximation one finds, up to logarithmic corrections,
 \begin{equation}
 \ln K \approx S \log (S) -\left(\frac{n_0}{2}+\frac{S}{2}\right) \log \left(\frac{n_0}{2}+\frac{S}{2}\right)-\left(\frac{S}{2}-\frac{n_0}{2}\right) \log \left(\frac{S}{2}-\frac{n_0}{2}\right)+\frac{1}{2} S \log (p q).
\end{equation}
This function has a maximum at
\begin{equation} \label{eqS}
S^* = \frac{n_0}{\sqrt{1-4 p q}},
\end{equation}
so the length of the most probable path scales linearly with $n_0$ up to a numerical factor that diverges at the transition point $p=q=1/2$ but otherwise is ${\cal O}(1)$. Note that our approximate expression counts all $S$ step pathes from $n_0$ to zero, including those that ``overshoot" and reach the negative region before returning to zero. The exact number of pathes of length $S$ is thus smaller than ${S \choose S/2+n_0/2 }$, meaning that Eq. (\ref{eqS}) overestimates the length of the typical trajectory. Since we have optimized the logarithm of the probability distribution function, the main contribution must come from the optimal path in the large $n_0$ limit.

This argument may be translated more or less directly to our system, that for low densities preforms a bias random walk along the log-abundance axis. It becomes exact if one neglects demographic stochasticity and impose density dependence by reflecting boundary conditions at $N$, the ceiling model used in \cite{lande2003stochastic} (see Section \ref{universality}), so the typical path to extinction is a sequence of ${\cal O} (\ln N)$ bad years.

\end{document}